# ARTICLE

## Above and beyond the laser–induced anti-Stokes broadband white light emission in rare–earth–manganese perovskites

Talita J. S. Ramos,*[a,b,c] Ágata Musiałek,[a] Robert Tomala,[a] and Wiesław Stręk[a]

We carried out a joint experimental and theoretical investigation of the laser-induced anti-Stokes white emission (LIWE) observed in $NdMnO_3$ nanocrystals. We evaluate numerically the mechanisms of the multiphoton ionisation, avalanche process, and the blackbody radiation, to provide a phenomenological model to interpret, and even predict, the LIWE phenomena. Through the theoretical modelling, photophysical and photoconductivity characterisations, we determine the ionisation rates (0.63—1.9 $s^{-1}$), ionisation fraction of the $Nd^{3+} + h\nu \rightarrow Nd^{4+} + e^-$, $Mn^{3+} + h\nu \rightarrow Mn^{4+} + e^-$, and $Mn^{2+} + h\nu \rightarrow Mn^{3+} + e^-$, from ionization energies ~0 until 52.58 eV, excitation at 808 and 975 nm, and different pressure conditions ($10^{-6}$—$10^3$ mbar). We propose a mathematical description of the LIWE generation as a dynamical process that explains the supralinear dependence of the integrated intensity of the emission with the laser power density of excitation, pressure, as well, the photophysical kinetics in order of ~ms to s, and the photocurrent (~nA) observed under excitation at high values of laser power density (~kW·$cm^{-2}$).

## Introduction

Laser–induced anti-Stokes broadband white light emission (LIWE), characterised by continuous spectrum in the visible region and a non-linear dependency of emission intensity on the excitation power, was first reported in 1994,[1] and nowadays observed in several optically active materials, encompassing organometallic clusters,[2] carbon-based materials,[3,4] hybrids,[5] and inorganic structures.[6,7] This phenomenon has been interpreted in terms of multistep photon ionisation by Costa 31 years ago,[8] and later as photon avalanche,[9] intervalence charge transfer,[10] blackbody radiation,[11,12] among others.[13,14] Due to the complex dynamics of the LIWE generation, its origin remains under debate; meanwhile, applications to this emission have emerged in the literature, including on-chip spectroscopy,[4] temperature/pressure sensing,[15,16] photovoltaic enhancement,[17,18] solid-state lighting,[14] and photothermic therapy at the nanoscale.[19,20]

The outstanding optical features typical of rare-earth ions,[21] namely, the quantum efficiencies of up to 100%, narrow emission bands, high colour purity degree, and long lifetimes, combined with the wide-ranging properties of perovskite-type oxides ($ABO_3$) as ferromagnetism,[22] ferroelectricity,[23,24] piezoelectricity,[25,26] and optics[27] suggest the rare-earth perovskites as multi-functional materials.

In this work, neodymium perovskite manganese ($NdMnO_3$) was studied for its structural and photophysical properties. Our contribution is a detailed interpretation of the LIWE in terms of sample composition, material compression, bandgap energy, several excitation conditions (wavelength, cycles, and excitation time), pressure, and temperature.

As a result of continuous wavelength (CW), is the laser excitation at 1.27–1.53 eV (~kW $cm^{-2}$) capable of promoting avalanche or multiphoton ionisation processes into materials that require ionisation energy as ~0.5 eV, 5 eV,[7] or even 50 eV[28,29]? In this work, we propose how the law of conservation of energy can be applied to describe the mechanism that generates the laser-induced white light emission characterised by continuous spectra from the visible and extended to the IR region, and analyse numerically the multistep photon ionisation and avalanche processes in rare-earth-manganese perovskite nanocrystals.

## Results and discussion

### Structural and morphological analysis

XRD analysis confirms the formation of the orthorhombic NdMnO₃ phase (ICSD No. 7221066), with no detectable secondary phases or impurities. The crystallite size, estimated using the Scherrer formula, corresponds to 136 ± 8 nm (see Table S1 with lattice parameters and unit cell volume). TEM images (Fig. S1) reveal semi-faceted to oval-shaped nanocrystals averaging 150 nm, as previously observed in the literature for LnMnO₃.[30]

### UV-Vis-NIR Characterisation

Room-temperature diffuse reflectance of NdMnO₃ consist of continuous spectra that can be explained due the overlap of the bands to the spin forbidden 5d transitions, described for charge transfer between the manganese and oxygen ions, in the octahedral crystal field, namely, $^4T_{1g}(4P)$, $^4E_g(4D)$, $^4T_{2g}(4D)$, $^4E_g(4G)$ and $^4A_{1g}(4G)$, $^4T_{2g}(4G)$ and $^4T_{1g}(4G)$, from the $^6A_{1g}(6S)$ ground state, identified at 310, 342, 363, 403, 444 and 540 nm, respectively.[31,32] As suggested by comparison with the international standard solar spectrum (ASTM G-173-03, ISO 9845-1), the NdMnO₃ shows a potential resonance spectral with commercial photovoltaic panels. The optical band gap (Eg) was calculated using the Schuster-Kubelka-Munk formula,[33–35] which considers a function F(R∞), where R∞ is the diffuse reflectance of the sample, as shown in Eq. 1:

$$F(R_\infty) = \frac{(1 - R_\infty)^2}{2R_\infty}$$   Eq. 1

The obtained value of Eg for NdMnO3 is 0.46 eV, as shown in Fig. 1, which fits the values reported in the literature.[36–40] The value to the Eg reported here is explained due to the strong hybridisation between Mn³⁺-3d and Nd³⁺-4f states, as suggested by the broadband spectra, as discussed before.[40]

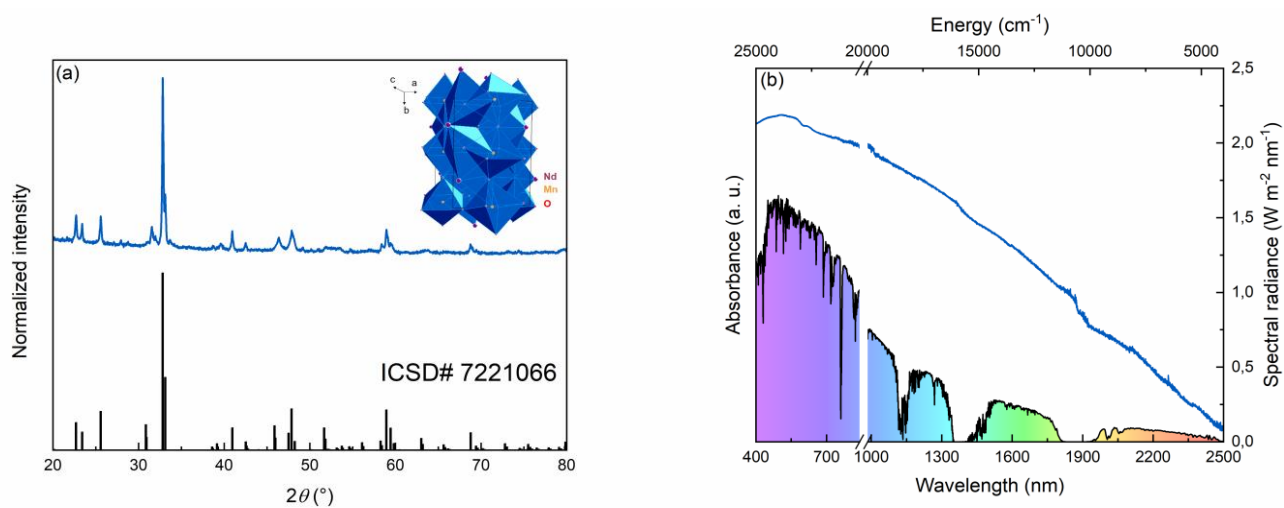

Fig. 1. (a) Diffraction pattern of NdMnO₃. Phase quantification was obtained using the software X'Pert HighScore Plus. Insert with the unitary cell view along [001]. (b) Absorbance of the samples NdMnO₃ compared to the international standard solar spectrum (ASTM G-173-03, ISO 9845-1).

### Laser-induced anti-Stokes broadband white emission (LIWE)

Laser-induced anti-Stokes broadband white emission (LIWE) was observed in NdMnO₃ under CW-λ$_{exc}$= 808 nm, or 975 nm, as a function of power density (P$_D$), pressure, and sample compression. The experimental methodology applied in the photoluminescence analysis and the data routine treatment are described in the SI and were based on well-established procedures reported in the literature.[15,41] This comparative analysis was obtained through the spectra acquired from nanocrystals and the bulk material. In all the studied cases, a non-structured broadband emission from the visible region to the NIR is observed, as shown in Fig.2 and Fig. S4–Fig. S9.

The well-known electronic structure of the Nd³⁺ presents resonance levels at 808 nm.[42] The laser excitation at this wavelength is usually ideal for exciting systems involving discrete levels of the neodymium ions. However, we noted that the excitation at 975 is much more efficient in promoting the LIWE. As can be seen in the emission spectra, the observed LIWE does not originate from 4f levels and/or due to the broadening of the typical intra-f transitions, as also indicated by the theoretical modelling presented further in this work. It is noteworthy that this emission, perceived by our eyes as white bright light, when converted to the chromaticity diagram, frequently corresponds to points in the orange-yellow region.[7,11,15]

The evaluation of colour purity of the emission characterised by unstructured broadband spectra, which extends over the visible range, is not relevant. In the present case, the emission spectra contain spectra components located outside the visible region that affect the intensity but do not influence the perceived colour. The objective of the CIE evaluation presented herein is to demonstrate the trend lines for the emission colour as a function of laser power density.

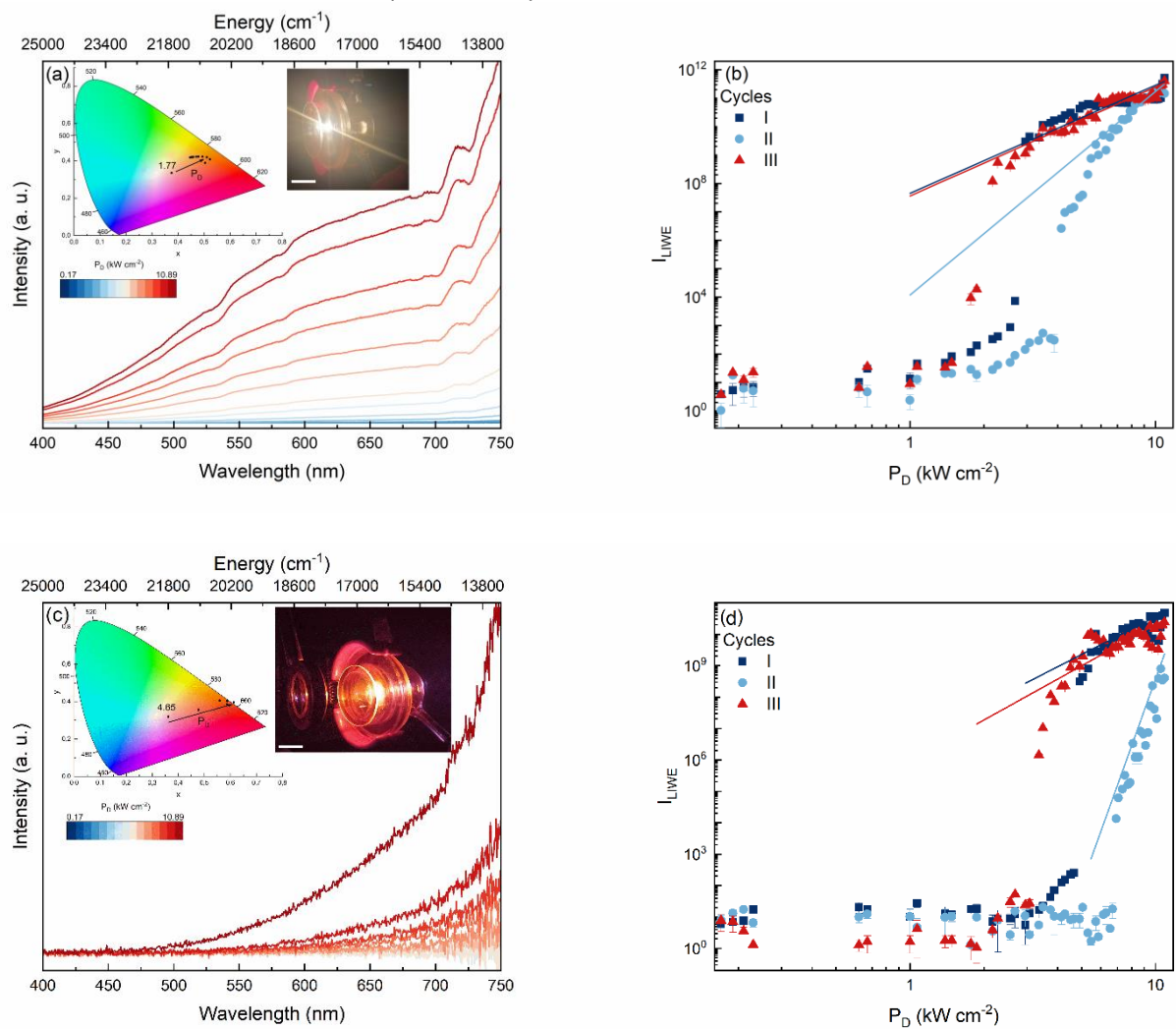

Fig. 2. (a) Photoluminescence spectra to NdMnO$_3$, obtained under CW-$\lambda_{ex}$=975 nm, analysed in the compacted form (ceramic) as a function of laser power density at (a) 10$^3$ mbar and (c) 10$^{-6}$ mbar, and their respective log–log plot of the power density (P$_D$, kW cm$^{-2}$) dependence of the integrate intensity of laser induced white light emission (I$_{LIWE}$) under different cycles of measurements. Cycles I and II correspond to the UC-emission spectra at a single sample spot under an increase and a decrease of the laser power density (P$_D$), respectively. Cycle III corresponds to the spectra collected at multiple points on the sample surface, each point corresponding to a P$_D$ value—inserts showing the CIE diagram of the obtained emission profile and photos recorded under 10.89 kW cm$^{-2}$.

## Power law to LIWE

The power law describes the number of photons (n) involved in the upconversion emission (I) originating from a quantised state and promoted by the excitation power density (P$_D$), through the slope obtained for the $I \propto P_D^n$.[43] The supra-linear dependence between the integrated emission intensity for the laser-induced broadband white light (I$_{LIWE}$) and P$_D$, is shown in Fig. 2 and Fig. S4–Fig. S9, to nanocrystals and to the bulk material, respectively. The calculated n ranged from 0.9 ± 0.3 to 21 ± 2, depending on the packing density degree of the sample, pressure, and, most notably, the excitation cycle. The values observed suggest that this relation cannot be interpreted as the

number of photons involved in the LIWE, as discussed previously.[11,15,41]

The multiphoton ionisation and avalanche processes are cited to explain the origins of the LIWE.[7–9,44,45] The multiphoton ionisation mechanism, proposed by Keldysh,[46] describes the emission promoted from the interaction of a pulsed laser beam (from 10$^{12}$–10$^{19}$ W cm$^{-2}$) via multiphoton absorption.[46–49] The similarities between the multiphoton ionisation and the laser-induced broadband white light emission concern the step-by-step behaviour observed for the power law and the emission favoured at low pressure. In the photon avalanche process, a sigmoidal relation between emission intensity and pumping power, associated with the high-order number of the photons involved in the emission, is identified in Ln-based systems,[50–53] and is attributed as the origin of the LIWE. However, those mechanisms do not explain the sample packing density effect, excitation cycle and pressure dependencies observed in the LIWE in several systems.[7–9,15,44,45] The $P_D$ applied (<10³ W cm⁻²) to promote the LIWE in the NdMnO₃ is too low to obtain the multiphoton excitation, as can be seen on the ionisation rates discussed here. As shown in Fig. 2 and Figs. S4–S9, the effect of sample compression and pressure is remarkable, indicating the influence beyond the laser excitation to generate the LIWE.

In the analysis of different cycles of laser irradiation, we observe the stability, uniformity, repeatability, and non-hysteresis profile of the white light emission as a function of laser power density. This analysis reveals a decrease in integrated intensity in the second cycle of excitation compared to the first and third cycles. The emission suppression after the laser irradiation suggests some modification on the emissive surface. We can see this effect by comparing the first with the third cycle (which generally shows higher integrated intensity). As indicated by the XRD and TEM analysis performed after the laser exposure, as presented in Fig. S17. We observed the annealing on the surfaces of the samples, which explains the saturation effect under laser irradiation and the increase in the value of the threshold in the power density to promote the laser–induced anti-Stokes broadband white light emission. The comparison between cycles I and III shows that the emission is not attributed to the degradation of the samples, as evidenced by the repeatability of the emission process.

**Temperature dependence on the LIWE**

To evaluate the eventual influence of the temperature on the mechanism that promotes the anti-Stokes broadband white light emission, we performed the adjustment of the emission spectra to the Planck curve distribution, as described in Eq. 2:

$$L_{BB} = \frac{A}{\lambda^{-n}}\left(e^{B(T)/\lambda} - 1\right)^{-1} \qquad \text{Eq. 2}$$

$$A = 2\pi hc^2, \; B(T) = hc/(k_B T)$$

Here $L_{BB}$ denotes the spectral radiance, which is proportional to the radiation intensity through the $A$ and $B(T)$ parameters, λ (λ) is the wavelength, h is Planck's constant, and c is the speed of light, as outlined above.[15,41] The exponents for the wavelength (n) were evaluated from 3.5 to 6.0. The best fits to the emission spectra (correct to spectral irradiance) were obtained for n equal to 5.0 and 3.5 for the nanocrystals and for the bulk materials, respectively. This observation is explained by blackbody radiation theory,[54–56] where the nanocrystals—due to their larger inter-particle distances—exhibit reduced energy dissipation. Therefore, the nanocrystals exhibit an optical behaviour similar to that of a blackbody, whereas the bulk behaves as a graybody.[41] As can be seen in Figs. 10–13, and Tables S4–S12, the emission spectra obtained for all studied systems are well-fitted to a Planck distribution curve. The eventual deviation from the theoretical curve is observed under ambient pressure and sample compression conditions. Those results highlight the contribution of non-radiative energy dissipation mechanisms, namely, convection and conduction, as described later in the energy balance equations rate.

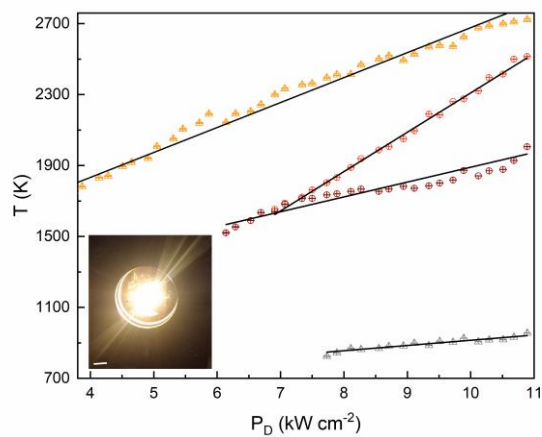

Fig. 3. Power laser density ($P_D$) dependence of the temperatures, fitted from emission spectra obtained under 808 nm (circles) and 975 nm (triangles) at ambient pressure (yellow and orange) and vacuum (grey and brown)—data available in the SI, Tables S5–S9. The lines are the best linear fits to the data (Table S10). Insert with photo of laser-induced white emission obtained under CW-excitation at 975nm, 10.89 kW cm⁻².

In a comparative study between the spectra obtained at different pressure conditions, it is frequently reported in the literature that the temperature associated with the emission spectra is higher for samples conditioned under vacuum.[13,15] We can explain this observation due to the heat dissipation mechanism (convection) promoted by the molecules present in the gas/atmosphere. At ambient pressure, the temperature of the emissive surface will be lower and, consequently, the radiative emission will be less intense, as predicted by Costa et al.[12] Interestingly, we observe the higher temperature in the samples in the compressed form, and at ambient temperature. This suggests the significant contribution of the conduction mechanism in feeding back the excitation process, as observed before.[9] The spectra of NdMnO₃ obtained at 808 nm (Fig. S10) show intangible emission that corresponds to a temperature below 750 K. This temperature is far from the typical temperatures associated with the blackbody radiation-type, which confirms the temperature dependence of laser-induced anti-Stokes broadband white light emission.

A critical factor in evaluating the influence of temperature on the promotion of LIWE is the physical dimensions of the emitting surface relative to the temperature measured by a thermocouple or a thermographic infrared camera.[13,44,45,57,58] However, the readings for these methods do not reflect the temperature on the emissive surface under laser spot irradiation. Instead, they show the average temperature across the entire sample, which leads to an underestimate of the temperature associated with the emission spectra and may cause researchers to misinterpret the phenomena.[13,44,45,57,58] In this work, the bulk consists of a ceramic in a circular shape with a diameter of 0.5 cm and has an area of 0.19 cm². The spot size of the laser beam used in the excitation is 0.0001 cm², as discussed in the SI. Therefore, the temperature calculated from the emission spectra corresponds specifically to the localised region directly under laser irradiation, which represents only 0.07% of the total sample area. Therefore, a temperature of 2000 K, determined spectrally, reflects just the temperature in the confocal centre of the irradiated spot ($\sim 10^{-4}$ cm$^2$) and should not be interpreted as the temperature of the bulk, *i.e.*, as the temperature measured by a thermocouple, as reported previously.[13,44]

Another method for estimating the temperature of an emitting surface is the use of a thermographic infrared camera. Thermal imaging systems estimate temperature based on the total radiative power collected over a surface, typically under the assumption of constant emissivity. In contrast, the emission spectroscopy yields a wavelength-resolved emissivity profile. According to the blackbody radiation theory, the emissivity depends on wavelength and temperature. Therefore, assuming the emissivity as a constant introduces systematic errors into temperature measurements obtained via thermal cameras.[13] An additional source of error arises from the spatial resolution of thermal cameras, once the spot size varies from a few millimetres to several centimetres, depending on the distance-to-spot ratio and the specific camera model. The A40 M ThermoVision camera from FLIR Systems, used previously by our group,[57,58] shows a spot size >10 times larger than the focal laser spot, which implies the underestimation of the temperature on the laser focal point.

As demonstrated in Fig. 3, the temperature calculated by the emission spectra is linear with the power density of the laser used in the excitation. The deviations from the trend line can be explained by thermodynamic principles: while conduction and convection processes tend to stabilise as the system approaches thermal equilibrium, the radiative heat transfer persists as a dynamic mechanism.[56,59]

### Temporal dynamics of the LIWE

The photophysical dynamics of laser-induced white light emission under NIR laser excitation, at fixed laser power density, are characterised by a rise time pattern that includes an approximately exponential increase, a stationary phase during laser irradiation, and an exponential decay.[13,15,41,60] The integration time used to acquire the spectra is frequently too broad to detect potential fluctuations in the dynamics of LIWE under CW-laser excitation.

The photophysical analysis methodology and data processing procedures are described in the Supporting Information and adhere to established literature protocols.[15,41] To promote a deeper analysis of the phenomenon kinetics, we used the minimum value of the integration time available in the spectrometer to detect any effect on the order of microseconds scale and the integrated intensity (with the error bar) under different cycles of measurements and acquired at different points of the sample surface. The temporal dynamics profiles of the LIWE are shown in Fig. S15. In the analysis of the LIWE dynamics, we summarise the following conclusions:

—The photophysical dynamics profiles are strongly dependent on the integration time used for spectral acquisition, pressure and compression of the sample, which suggests more than one physical process associated with the LIWE generation.

—To the spectra obtained at the minimum power density to generate the emission (6.69 KW cm$^{-2}$), the rise time is approximately double when we apply higher values of maximum $P_D$ (10.89 KW cm$^{-2}$). In this case, the integrated intensities are smaller (as evidenced in the error bars) and are explained by the signal/noise ratio for the acquired emission profiles.

—The dynamics observed in the order of temporal scale from milliseconds to seconds, depending on the power dependency and pressure, corroborate that the LIWE do not generate due to the excitation or broadening of ion transitions, as indicated by the photoluminescence acquired on the visible and IR analysis reported here. The values observed (rise time $\sim 0.5 \pm 0.1$ s, decay time $\sim 0.26 \pm 0.07$ s) fit in the interpretation of the laser-induced white light emission as a thermal or blackbody-type radiation, as discussed previously.[15]

Indeed, the LIWE involves non-quantised heat transfer processes, unlike discrete upconversion emissions observed in lanthanide-based systems, which arise from electronic transitions between quantised energy levels. These processes show dynamics on a microsecond scale and generate white light emission as a result of the superposition of emissions at multiple wavelengths.[61,62]

### NIR analysis to LIWE

The emission spectra collected in the near-infrared (NIR) region under 808 nm excitation closely resemble those obtained under 975 nm excitation (see Fig. 4, Fig. S4–Fig. S9), and notably do not correspond to the characteristic 4f-emissions of Nd$^{3+}$ or even the Mn-ions.[63,64] The present analysis of NIR spectra reveals the effects of higher temperatures on the sample surface and the influence of pressure and sample packing density on the LIWE process.

As shown in Fig. 4, the bulk material analysed under ambient pressure exhibits a spectral tail from $\sim 1115$ nm to 1915 nm, which aligns with the profile of a blackbody radiation curve. The observed intensities increase proportionally with the laser power density, indicating that this tail arises from a light-to-heat conversion process induced by the excitation. Consistent with the visible-range spectral analysis, once a power density threshold is surpassed—corresponding to elevated local temperatures—a broadband emission in the NIR emerges. This emission is attributed to thermal excitation (thermionic excitation) of a fraction of Nd$^{3+}$ ions, in agreement with previously reported behaviour for lanthanide oxides.[11] This thermally induced broadband emission is particularly evident in the emission observed for the nanocrystals at reduced pressure ($10^{-6}$ mbar), where limited conduction and convection lead to an increase in local emission temperature. For bulk samples under low pressure, the convection is suppressed, and the prominent conduction process leads to a weaker broadband emission due to the heat dissipation. Overall, the observed dependence of emission behaviour on pressure and the sample's packing density underscores the essential roles of conduction and convection in dissipating the laser-absorbed energy. Those observations are in line with the mechanisms described by the rate balance equations discussed in this work.

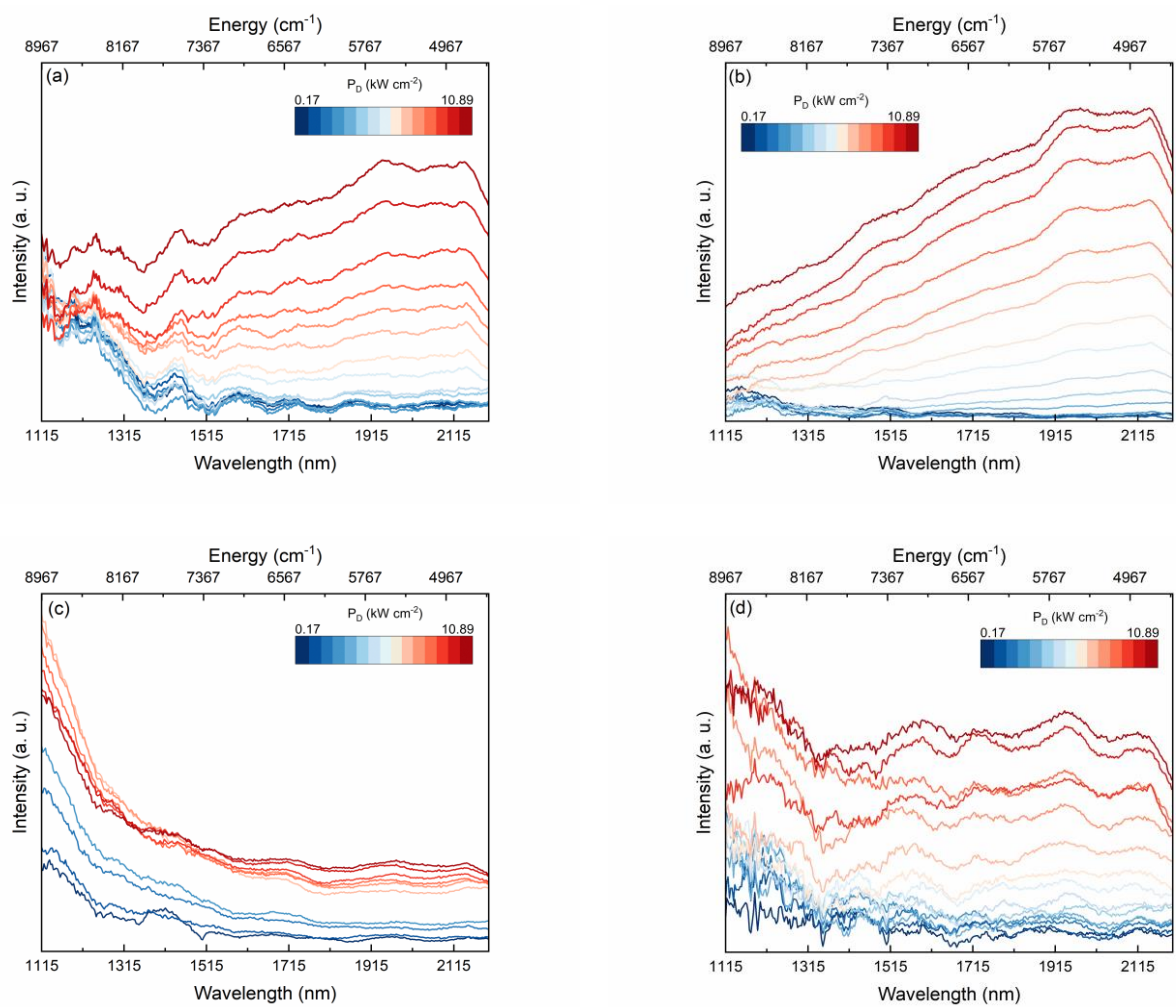

Fig. 4. (a) NIR photoluminescence spectra of NdMnO₃, obtained under CW-λₑₓ=975 nm, analysed as nanocrystals at (a) 10³ mbar and (b) 10⁻⁶ mbar, and in compacted form (ceramic) at (c) 10³ mbar and (d) 10⁻⁶ mbar, as a function of laser power density.

## Modelling the LIWE

A single atom can absorb photons for which the probability that the atom will be induced to move into an excited state is determined by the inverse of the rate of stimulated transitions between the two states of the absorption transition.[65] As well known by the photon ionisation mechanism, the photoionisation rate ($P_{photo}$) is determined by the product between the optical cross-section of the ion ($\sigma_{opt}$) and the photon flux of the excitation source ($\Phi_{excitation}$), both at specific wavelength. As reported previously,[66–68] the optical cross-section of the Nd³⁺ ion is $3.3 \times 10^{-20} m^2$ at 808 nm, and $1.0 \times 10^{-20} cm^2$ at 975 nm. The photoionisation rate is calculated by Eq. 3:

$$P_{Photo} = \sigma_{opt} \; x \; \phi_{excitation} \qquad \text{Eq. 3}$$

The photon flux of the lasers MDL-H-808-6W and MDL-H-975-5W, as provided by the manufacturer, corresponds to $6.0 \times 10^{23} m^{-2}s^{-1}$ and $6.25 \times 10^{23} m^{-2}s^{-1}$, therefore, the photoionisation rates are

1.9 $s^{-1}$ and 0.63 $s^{-1}$, respectively. The low values of the $P_{photo}$, in comparison with the observed spectra, indicate that the multiphoton ionisation does not provide a mechanistic explanation for the LIWE observed. The ionisation fraction, described previously,[69,70] and presented by Eq. 4, was calculated considering the electron number density ($n_e$) of a solid system and the excitation conditions (pressure, $\lambda_{ex}$ and $P_D$). Firstly, we determine the $n_e$ generated from the Nd³⁺, and also for Mn³⁺/Mn⁴⁺and Mn²⁺/Mn³⁺, into the NdMnO₃ (detailed $n_e$ calculation on the Supplementary Information):

$$\frac{N_{i+1}}{N_i} = \frac{2Z_{i+1}}{n_e Z_i}\left(\frac{2\pi m_e k_B T}{h^2}\right)e^{-E_{ion}/k_B T} \qquad \text{Eq. 4}$$

Where:

$N_{i+1}$ and $N_i$ are the number densities of ions N in ionisation states i+1 and i,

$Z_{i+1}$ and $Z_i$ are the partition functions for degenerate states of $Nd^{3+}$ and $Nd^{4+}$, then, assuming that the partition functions for the two ionisation states are nearly equal, i.e. $Z_{i+1} \approx Z_i \approx 1$.

$m_e$ is the electron mass ($9.109 \times 10^{-31}$ kg),

$k_B$ is the Boltzmann constant ($8.617 \times 10^{-5}$ eV $K^{-1}$),

h is Planck's constant ($6.626 \times 10^{-34}$ J·s),

$E_{ion}$ is the ionisation energy required to ionise from state i to i+1,

T is the temperature (in K) and is related to the laser power density of the laser excitation.

We calculate the ionisation rates that $Nd^{3+} + h\nu \rightarrow Nd^{4+} + e^-$ considering the bandgap of the $NdMnO_3$ as 4.60 eV, as determined previously.[40] We evaluate the possibility of $Nd^{3+}$ is located in a defect state below the conduction band, and also using the Keldysh approximation[47] (where $E_{ion} \sim Eg = 0.46$ eV), for the $NdMnO_3$ irradiated at CW-808 and 975 nm at ambient pressure and vacuum, as can be seen in the Fig.5a and Fig.S16. We performed calculations of ionisation fractions for the manganite ions, specifically $Mn^{3+} + h\nu \rightarrow Mn^{4+} + e^-$, and $Mn^{2+} + h\nu \rightarrow Mn^{3+} + e^-$, as shown in Fig. 5b and 5c, respectively. The ionisation rates determined from $10^{-209}$ up to $10^{-31}$ reveal that the fraction of $Nd^{3+}$ ions that ionise to $Nd^{4+}$, or even the pairs $Mn^{3+}/Mn^{4+}$, and $Mn^{2+}/Mn^{3+}$, are extremely small, effectively zero, even at higher values of laser power density (kW $cm^{-2}$). The ionisation fraction determined in this work indicates the mechanism of multiphoton ionisation, proposed by Keldysh,[46,47] or the multistep ionisation suggested by Costa,[8] does not explain the origin of laser-induced broadband white light emission obtained under excitation power density < 10.89 kW $cm^{-2}$. Aiming to clarify the LIWE phenomenon, we propose the phenomenological model detailed below..

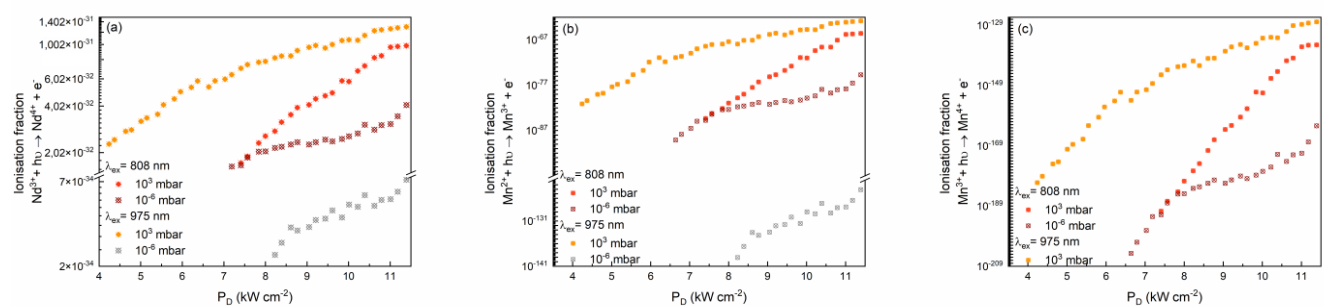

Fig. 5. Ionisation fraction to (a) $Nd^{3+} + h\nu \rightarrow Nd^{4+} + e^-$, (b) $Mn^{3+} + h\nu \rightarrow Mn^{4+} + e^-$, and (c) $Mn^{2+} + h\nu \rightarrow Mn^{4+} + e^-$, as a function of laser power density of excitation (0.17-10.89 kW$cm^{-2}$) at CW-$\lambda_{ex}$= 808 nm, or 975 nm, and different pressure conditions, and $E_{ion} \sim 0.46$ eV.

## Phenomenological model of the LIWE

We propose a mathematical description of the interaction of radiation with a nanoparticle based on the general power balance equation, as supported by multiple publications,[12,41,71–75]; however, this description has not been combined and evidenced in the experimental results, as presented in this work by Eq. 5.

$$W_{absorption}^{laser} = W_{Radiation} + W_{Conduction} + W_{Convection} \quad \text{Eq. 5}$$

Where, an effective absorption cross-section can describe the absorption rate, so the rate of energy absorption well explored in the literature[12,74,75] with $\sigma_{abs}$ (in $cm^2$) being the effective absorption cross-section of the particle at a given frequency v, or wavelength λ, and $P_D$ (in W $cm^{-2}$) is the power density of the excitation source (e.g., a continuous laser beam).

The radiative emission rate is described mathematically according to the type of emission identified. For systems with lanthanide ions, in which discrete and well-defined peaks are identified, the Judd-Ofelt-Malta theory[76–79] presents the corresponding operators for this emission. In this work, the emission profile presents only the thermal radiation represented by the Planck equation; therefore, the radiative rate corresponds to Eq. 6, where $W_{emission}^{bb}$ corresponds to the spectral radiance.

$$W_{Radiation} = W_{emission}^{bb} \alpha T^n \quad \text{Eq. 6}$$

We performed the adjustment of the photoluminescence profiles with the Planck's distribution curve, varying the exponential (indicated here by n) with the wavelength, and best fits were obtained with n=3.5 to the ceramic and 5 to the powdered samples, which corroborates the proposed by blackbody radiation theory[54]

Using the balance rate proposed here, we can explain the photophysical behaviour observed in the nanocrystals and bulk samples, under several excitation conditions ($P_D$, $\lambda_{ex}$, and pressure), see Fig. S13a and Fig. S13b, respectively. In the case of the bulk samples, the contribution of $W_{conduction}$ is more pronounced due to higher contact between the particles, resulting in a decrease in the $W_{Radiation}$ and a shift to higher values of the power density threshold to promote LIWE ($P_{D, threshold}$). In the same way, it is possible to understand the changes in the PL profile between the samples at ambient and at vacuum. At low pressure, we observe the domain of the $W_{Radiation}$ contribution and the decrease of the $P_{D,\ threshold}$. The deviation from the fit of the emission spectra to the Planck distribution curve, considering the spectra obtained at the ambient pressure (Fig. S12 and Fig. S13), evidenced the role of dissipation of energy by non-radiative pathways (namely, $W_{conduction}$ and $W_{convection}$) for the $W_{radiative}$ parcel. The prediction character of the balance rate proposed here is based on the absorption process, which is determined by the cross-section on the wavelength used for excitation, the component of thermal conduction, evaluated by thermal characterisation of the samples, and the convection parcel that can be minimised under vacuum conditions.

### Photoconductivity analysis of LIWE

Measurements of the photoconductivity within $NdMnO_3$ nanocrystals were performed in the bulk samples using high-resistance electrometry under laser excitation at room temperature and vacuum conditions ($10^{-6}$ mbar), through the methodology developed and extensively explored in our group in previous publications.[3,60,80] The resistance measurements were performed with a probing voltage of 10-150 V. As shown in the PL spectra, the LIWE emission to $NdMnO_3$ is promoted just above the power threshold of optical power. For excitation below this threshold, no photoconductive effect is observed.

In the photoconductive analysis, we observe the resistivity oscillation in just one order of magnitude and time scale of seconds. As detailed in the SI, those results indicate that the multiphoton ionisation mechanism, avalanche processes, charge transfer, and oxygen vacancies do not take place in the generation of the photoconductivity associated with the laser–induced anti-Stokes broadband white light emission for the $NdMnO_3$, and for several systems reported in literature.[3,45,60,80,81]

The decrease in the resistivity, observed during the LIWE generation for the rare–earth–manganese perovskites and for the previous works,[3,45,60,80,81] fits in the interpretation of the decrease in the resistance of a material associated with the substantial increase in the temperature. This relation was described by Richardson–Dushman, as shown in Eq.7:

$$J_{Richardson-Dushmann} = AT^2 exp\left(\frac{\phi}{k_BT}\right) \qquad \text{Eq. 7}$$

Where J indicates the current density, T is the temperature of the emissive surface, $k_B$ is the Boltzmann constant, and $\phi$ corresponds to the work function (fixed and typical for the material analysed). The error associated with J is presented in the SI. We modulate the current density taking into account the approximation, valid to the multiphoton ionisation emission, of the $\phi \sim E_g$ from 0.46 to 4.6 eV, as outlined above [36–40]

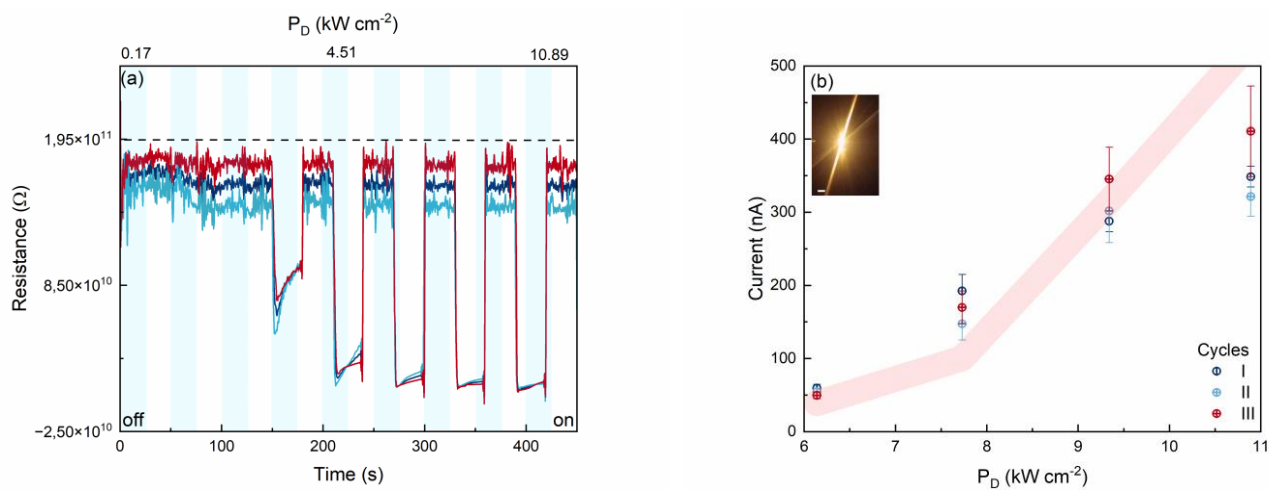

Fig. 6. (a) LIWE photoresistance of $NdMnO_3$, analysed as a bulk material, with 30 s of on/off cycles of laser excitation at 975 nm. Note to the saturation regime of the photoconductivity reached around 300 s. (b) LIWE photocurrent generation as a function of the laser power density and the tendency line to the current density calculated using the temperature determined from the emission spectra (Eq. 6). Insertion with an emissive surface during laser excitation at 975nm, 10.89 kWcm$^{-2}$.

The laser–induced anti-Stokes broadband white light emission is a surface phenomenon; therefore, the photocurrent associated with it must be normalised by the conductive area and presented in terms of current density to be compared with values referenced in the literature. Our group uses the exact dimensions for preparing ceramics, allowing the data provided to be directly compared. Frequently, an oscillation of resistivity of up to $10^7 \rho$, associated with the photocurrents in the order of nA to a maximum of µA, during the LIWE generation, is observed on a time scale of seconds.[82–84] These results can be explained by the decrease in resistivity with the increase in temperature on the sample surface under enhanced laser power density irradiation. The tendency line observed in Fig.6 shows the current density calculated for the temperature indicated by the emission spectra obtained at 808 and 975 nm. It shows the direct relation between the LIWE and the temperature effect induced by the excitation sources.

The XRD and TEM analyses of samples subjected to multiple cycles of CW-laser excitation (Fig. S17) show a broadening of the XRD peaks and an increase in the average particle size. These results provide clear evidence of the annealing process induced by NIR irradiation and support the dominant role of the temperature in the laser–induced anti-Stokes broadband white light emission discussed in this work.

# Experimental

### Synthetic Procedures

The samples were synthesised by self-combustion using a mixture of $RE(NO_3)_3$, where RE=Nd or Yb, $Mn(NO_3)_3$, and urea (ratio 1:2.5), subjected to a furnace at 10°C min-1, 650 °C for 5 minutes, followed by calcination at 1100 °C for 6 hours.

### Characterisations Methodologies

The XRD patterns were collected at room temperature between 10 and 89 degrees (in 2θ) by an X'PERT PRO (PANalytical, The Netherlands) diffractometer equipped with a PIXcel ultrafast line detector and Soller slits for CuKa1 radiation (1.5406 Å, step: step 0.36°min-1). The X-ray tube settings were 30 mA and 40 kV. TEM images were performed using a Philips CM−20 SuperTwin TEM microscope with a resolution of 0.25 nm and operating at 160 kV.

Absorption spectra were measured in reflection mode using an Agilent Cary 5000 UV-Vis-NIR spectrophotometer with an integrating sphere (Labsphere, 150 mm diameter, Spectralon-coated). A deuterium lamp (185−350 nm) and tungsten-halogen lamp (350−3300 nm) served as excitation sources. Detection was achieved with an R928 photomultiplier tube for UV-Vis (185−900 nm).

Photoluminescence spectra were collected under CW-808 nm excitation (MDL-H-808-6W, 808 ± 5 nm, $6 \times 10^{23}$ photons $m^{-2} s^{-1}$) or CW-975 nm excitation (MDL-H-975-5W, 975 ± 10 nm, $6.25 \times 10^{23}$ photons $m^{-2} s^{-1}$) in the visible range using the Ocean Optics QE Pro and in the infrared range, the NIRQuest 512−2.5, both from Ocean Optics. The filters used here were FESH750 or FESH900 in the visible detected range and FEL1100 in the infrared range from Thorlabs. The optical properties of the materials were investigated at ambient pressure and in a vacuum ($10^{-6}$ mbar obtained with the Turbomolecular Drag Pump TMH071P electronic drive unit TC 600, Pfeiffer). The power density measurements were performed using a Thorlabs NRT100 motorised linear translation stage to precisely control the distance between the sample and the confocal lens.
All of the collected data was corrected by the function of the filters and the detector's sensitivity. The experimental setup used for spectrum acquisition and all steps in data treatment are discussed in detail in the SI.

Photoconductivity was studied using high-resistance electrometry (Keithley 2400) under 808 or 975 nm CW-laser diodes in one-minute cycles of excitation and from 0.17 to 10.89 W cm$^{-2}$. All measurements were performed on the samples on a ceramic conditioner at $10^{-6}$ mbar using a vacuum cell supplied with a Turbomolecular Drag Pump TMH071 at room temperature. The ceramic, with a diameter of 5 mm and a thickness of 0.7 mm, was sintered at 0.200 g of the powdered sample (nanocrystals) compressed at 8GPa for 1 minute and 500 °C. Due to the compression, the ceramic corresponds to the bulk material composed of nanocrystals.

# Conclusions

Herein, lead-free perovskite $NdMnO_3$ was synthesised by a combustion reaction, producing crystals with high purity and a narrow average crystallite size distribution (136 ± 8 nm). Under continuous wavelength photoexcitation at 808 or 975nm, $NdMnO_3$ exhibits a transition from a non-emissive to a highly emissive regime, driven by power densities ($P_D$) up to 10.89 kW cm$^{-2}$ in nanocrystals and bulk material. The hypothesis that laser-induced broadband white light emission (LIWE) is promoted due to the excitation of the 4f-Nd$^{3+}$ levels, or excitation of manganite ions, is not verified in the results. The LIWE observed as an efficient warm to bright white light in the visible region, also in the NIR, stems from blackbody-type radiation.

The stable, repeatable, and tunable multi-coloured emission at a targeted excitation condition (wavelength, power density and pressure) shows potential applications in white light-emitting diodes development. The radiance profile obtained is very similar to the sun, with the spectral resonance regions of the commercial solar cells, indicating paths for applications to the $NdMnO_3$ beyond the thermal sensing presented here. The analysis of the NIR spectra discussed demonstrates the consequences of the increase in temperature at the material surface and highlights the influence of the pressure and the sample's packing density on the LIWE generation.

We evaluate mathematically the multiphoton ionisation mechanism and avalanche process, presenting the ionisation rates (1.9 s$^{-1}$ at $\lambda_{ex}$=808 nm and 0.63s$^{-1}$ at $\lambda_{ex}$=975 nm), and the ionisation fractions to the Nd$^{3+}$, Mn$^{3+}$ and Mn$^{2+}$ ions (from $10^{-209}$ up to $10^{-31}$). The results demonstrate that the multiphoton ionisation and the avalanche process do not take place in the origin of the LIWE obtained at power density from 0.17 to 10.89 kW cm$^{-2}$, under excitation at 1.27-1.53 eV. This study provides new insights into the physical mechanism that generates the laser-induced white light emission through a balance equations rate, in accordance with the photophysical behaviour, temporal dynamics observed in a time scale of milliseconds to seconds, and photoconductivity characterised by variation of resistance of 10-10$^7$ ohms and photocurrent of ~nA to mA.

## Author contributions

Talita J. S. Ramos performed morphological, photophysical, and photoconductivity characterisations, data analysis, theoretical modelling, and writing. Ágata Musialek synthesised the samples. Robert Tomala carried out the structural phase quantification and funding acquisition. Wiesław Stręk provides project administration and funding acquisition. All authors have approved the final version of the manuscript.

## Keywords

## Conflicts of interest

There are no conflicts to declare.

## Data availability

The data supporting this article have been included in the main manuscript and the Supplementary Information.

## Acknowledgements

This work was supported by the NCS project Grant No. NCN−2020/37/B/ST5/02399. The authors thank Dr. Karolina Ledwa (INTiBS) for the TEM analysis and Dr. Paweł Głuchowski (Graphene Energy) for preparing the ceramics analysed in this work. Talita J. S. Ramos acknowledges Prof. Dr. Ricardo Luiz Longo (UFPE) for suggesting the interpretation of LIWE using equilibrium rate equations and the fellowship from the Next Generation EU project P2022ALSMP VISIO (CUP B53D23025290001).

# Above and beyond the laser–induced anti-Stokes broadband white light emission in rare–earth–manganese perovskites

Talita Ramos,[a, b,c ,*] Ágata Musialek,[a] Robert Tomala[a] and Wiesław Stręk[a]

[a]Institute of Low Temperature and Structure Research, Polish Academy of Science, Okólna 2, 50-422 Wroclaw, Poland.

[b]European Laboratory for Non Linear Spectroscopy (LENS), via Nello Carrara 1, 50019 Sesto Fiorentino, Italy.

[c]Istituto Nazionale di Ottica, Largo Fermi 6, 50125 Firenze, Italy

*talita.ramos@lens.unifi.it

## Table of Contents

### 1. Structural and morphological characterisations

**Table S1.** Structure parameters for NdMnO₃ nanocrystals. Average crystalline size calculated by the Scherrer equation.

| Symmetry | Orthorhombic |
|---|---|
| Space group | Pnma (62) |
| a (Å) | 5.7881 |
| b (Å) | 7.5564 |
| c (Å) | 5.4101, c/2=2.705 |
| V (Å³) | 236.62 |
| Crystalline size (nm) | 136 ± 8 |

Structurally, $NdMnO_3$ has a distorted perovskite-type lattice with Mn-O octahedra sharing apical oxygen atoms, featuring ~2.81 Å Mn-O bond lengths in the ab plane, which explains the plan seen in HRTEM, Fig. S1a. As shown in Fig. S1b and Fig. S16b, we performed the structural and morphological characterisations before and after laser irradiation, respectively.

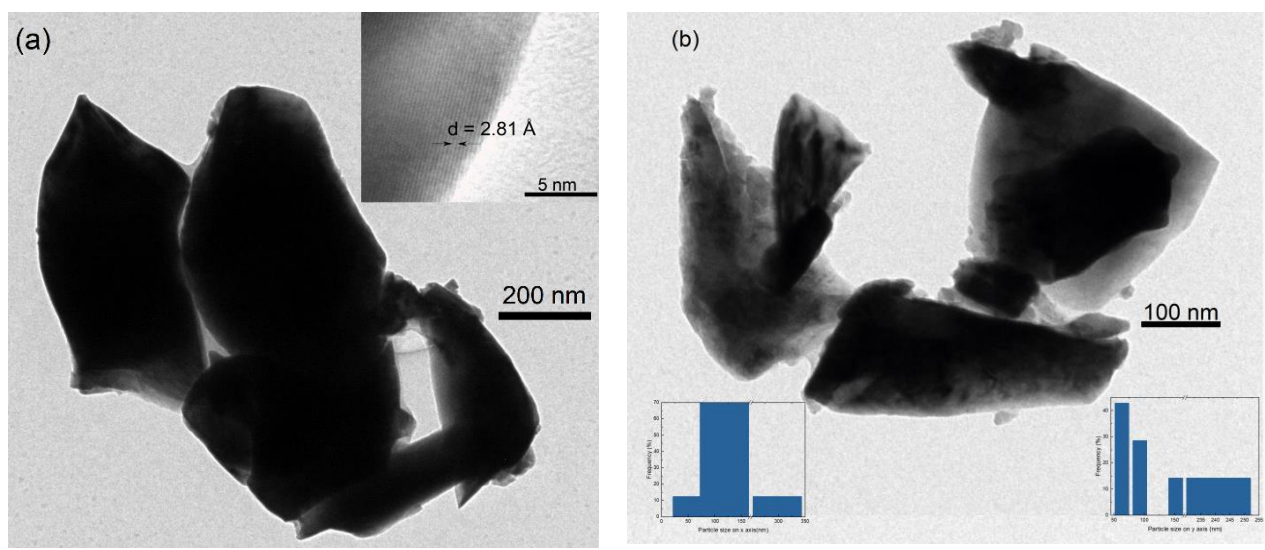

Fig. S1. TEM images of NdMnO₃, before laser exposure, at (a)200 nm and (b)100 nm of magnification. Insets in the figures show the HRTEM image and aggregate the histogram to the image on (b).

## 2. UV-Vis-NIR analysis

Room-temperature diffuse reflectance of NdMnO₃ consist of continuous spectra that can be explained due the overlap of the bands to the spin forbidden 5d transitions, described for charge transfer between the manganese and oxygen ions, in the octahedral crystal field, namely, $^4T_{1g}(4P)$, $^4E_g(4D)$, $^4T_{2g}(4D)$, $^4E_g(4G)$ and $^4A_{1g}(4G)$, $^4T_{2g}(4G)$ and $^4T_{1g}(4G)$, from the $^6A_{1g}(6S)$ ground state, identified at 310, 342, 363, 403, 444 and 540 nm, respectively.[1,2] As suggested by comparison with the international standard solar spectrum (ASTM G-173-03, ISO 9845-1), the NdMnO₃ shows a potential resonance spectrum with commercial photovoltaic panels. The optical band gap (Eg) was calculated using the Schuster-Kubelka-Munk formula,[3–5] which considers a function F(R∞), where R∞ is the diffuse reflectance of the sample, as shown in Eq. S1:

$$F(R_\infty) = \frac{(1 - R_\infty)^2}{2R_\infty}$$

Eq. S1

The obtained value of Eg for NdMnO3 is 0.46 eV, as shown in Fig. S2, which fits the values reported in the literature from ~0 to 4.8 eV.[6–10] The low value of the Eg reported here is explained by the strong hybridisation between Mn³⁺-3d and Nd³⁺-4f states, as suggested by the broadband spectra, as discussed before.[10]

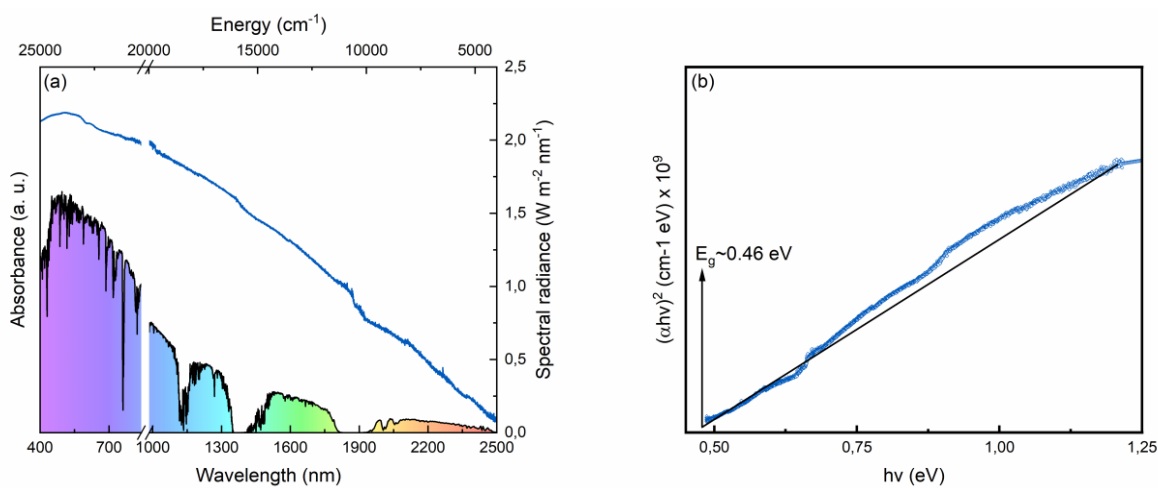

Fig S2. (a) Absorbance of the samples NdMnO₃ compared to the international standard solar spectrum (ASTM G-173-

03, ISO 9845-1)—optical bandgap determination using $(\alpha h \upsilon)^2$ versus $h\upsilon$ plot.

### 3. Experimental setup used to study the anti-Stokes laser-induced white emission (LIWE)

For the comparative analysis between different experiments, we report the excitation in terms of laser power densities. The laser spot area at the focal plane was determined following the methodology described by Saleh ([11] where the focused spot diameter ($W_0^{'}$), as given by Eq. S2, where f is the focal path of the lens, λ is the laser wavelength, and D is the incident beam diameter.

$$W_0^{'} = \frac{2f\lambda}{\pi D}$$
Eq. S2

$$A = \pi \left(\frac{W_0^{'}}{2}\right)^2$$
Eq. S3

The corresponding spot area was then calculated based on this value, as shown in the equation. S3. For intermediate positions along the beam path, spot diameters were estimated using a conical beam propagation model, as previously detailed in our works.[12,13] The experimental limitations, e.g. the range of detection to visible from 400 until 750 nm, and IR from 1013 to 2300 nm (see user's manual to respective detectors)[14,15] and showed in the Fig. S3, the need of use different filters, namely, FESH750/FESH900 and FESH1100, and the spatial orientation of the optical fibre to collect the signal at nearly 45° with the excitation laser beam (to avoid secondary effects like reflection and so on), the spectra acquisition in the visible range concurrent with the IR is not possible. Two different detectors have been used to collect the results on the visible and in the IR, so direct comparisons between the intensities in both regions are not possible.

It is important to note that due to the known variation in detector sensitivity based on the operating range (particularly for Si-based detectors and InGaAs-based IR detectors), the spectrum correction is essential for obtaining accurate spectrometer measurements, as highlighted in several previously published references.[16–22] All of the collected data was corrected by the function of the filters and the detector's sensitivity. The experimental setup used for spectral acquisition is shown in Fig. S3.

In the context of data processing, to detect the most minor changes in the spectra acquired as a function of intensity (photons per second) versus wavelength, the spectral conversion is performed using the Jacobian transformation, which presents the intensity in arbitrary units versus energy, as shown in Eq—S4 for the abscissa and Eq. 5 for the ordinate values. We analysed the data in terms of energy and wavelength; however, in order to facilitate comparisons with the results reported in the literature, the data were presented in terms of wavelength. Spectral conversion using the Jacobian factor is crucial when studying materials that have a wide spectral emission window, such as materials with blackbody-type emission. It is essential for extracting population information regarding the levels and areas related to the radioactive decays, as discussed previously.[23–25]

$$E = \frac{hc}{\lambda}$$
Eq. S4

$$f(E) = -f(\lambda)\frac{hc}{E^2}$$
Eq. S5

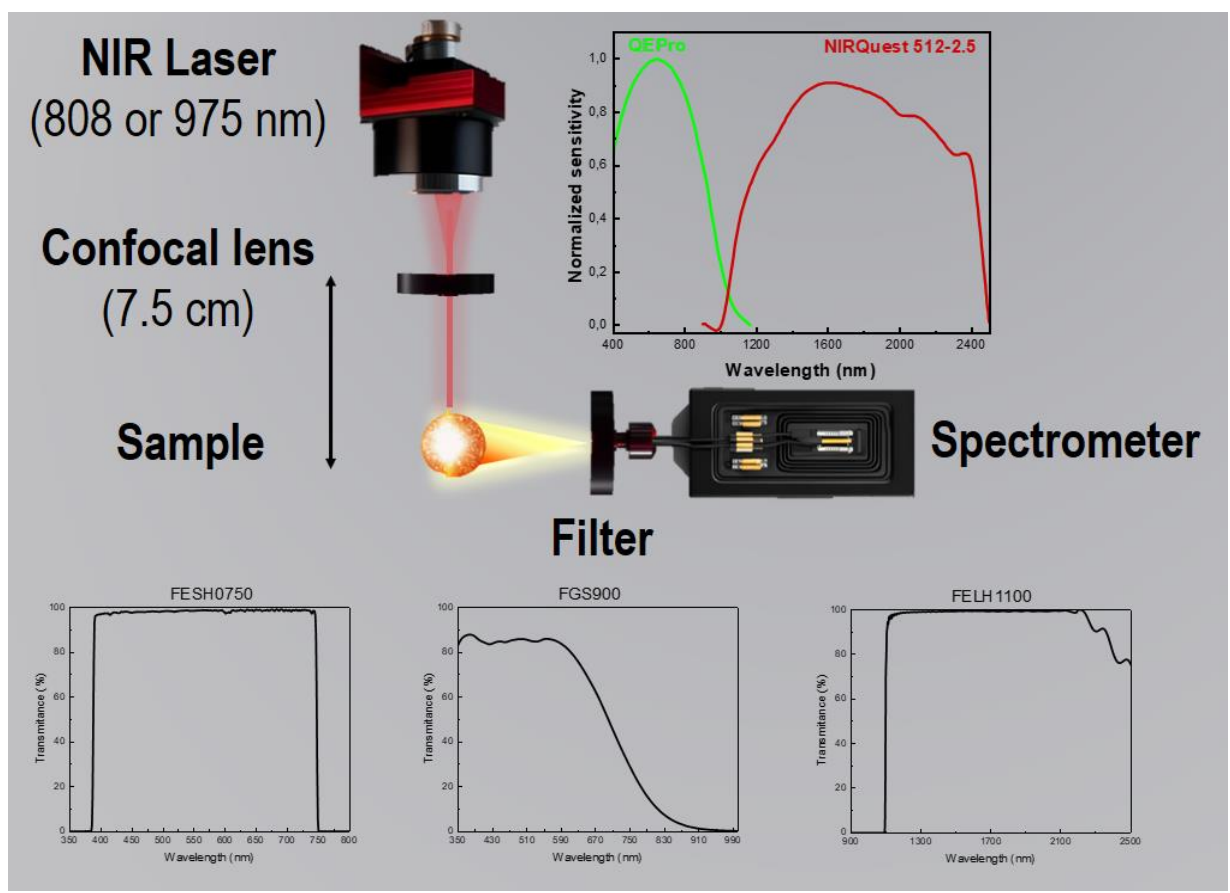

Fig S3. Illustration of the experimental setup used to study the anti-Stokes laser-induced broadband white emission (LIWE). Insert with the transmittance spectra of the filters (provided by Thorlabs) and the sensitivity profile (provided by Ocean Optics) to the spectrometers used, QE Pro and NIRQuest 512-2.5, in the visible and IR range, respectively.

### 4. Anti-Stokes laser-induced white emission (LIWE) at laser excitation in 808 nm: sample compactation and pressure dependencies

Fig. S4 and Fig. S5 show the PL to the bulk material, at ambient pressure and vacuum, respectively. The spectra obtained for the bulk at ambient pressure are shown in Fig. S6, and those obtained at vacuum are shown in Fig. S7. The analysis of the visible range does not show significant differences.

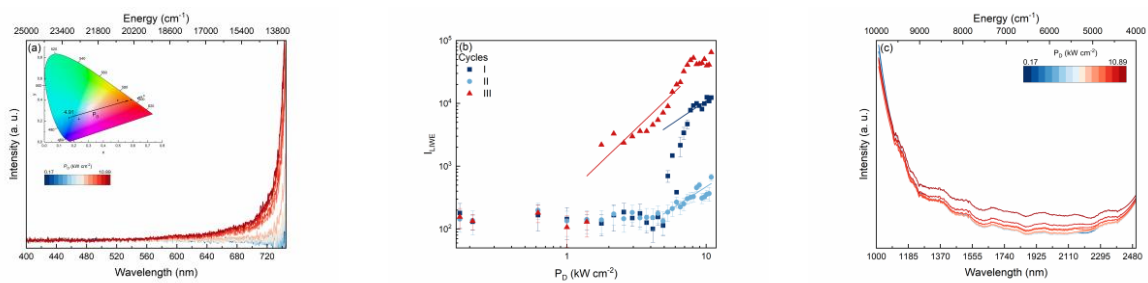

Fig. S4. (a) Photoluminescence spectra of NdMnO$_3$ nanocrystals at $\lambda_{ex}$=808 nm, 10$^3$ mbar and ambient temperature, in function of laser power density (P$_D$) in (a) visible and (c) IR regions. In the IR spectra, note the attenuation effect of the FEL1100 (Thorlabs). (b) Log–log plot of the power density (P$_D$, kW cm$^{-2}$) dependence of the laser-induced white light emission (I$_{LIWE}$) under different cycles of measurements. Cycles I and II correspond to the UC-emission spectra at a single sample spot under increasing and decreasing the laser power density (P$_D$), respectively. Cycle III corresponds to the spectra collected at multiple points on the sample, one point to each P$_D$ value—inserts showing the CIE diagram of the obtained emission profile.

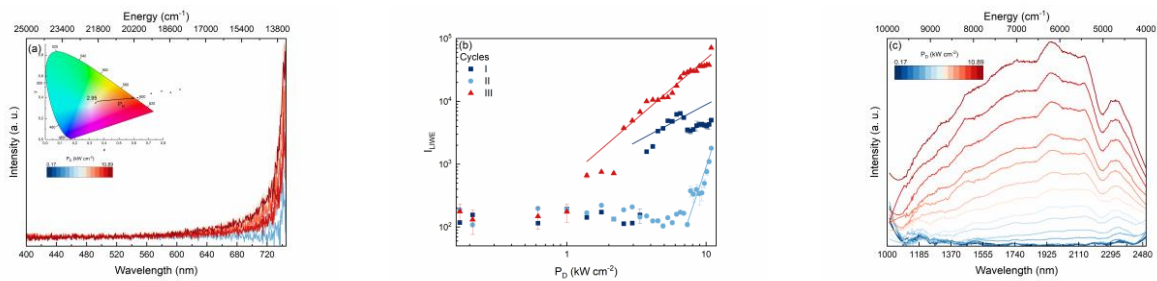

Fig. S5. (a) Photoluminescence spectra of NdMnO$_3$ nanocrystals at $\lambda_{ex}$=808 nm, 10$^{-6}$ mbar and ambient temperature, in function of laser power density (P$_D$) in (a) visible and (c) IR regions. In the IR spectra, note the attenuation effect of the FEL1100 (Thorlabs). (b) Log–log plot of the power density (P$_D$, kW cm$^{-2}$) dependence of the laser-induced white light emission (I$_{LIWE}$) under different cycles of measurements. Cycles I and II correspond to the UC-emission spectra at a single sample spot under increasing and decreasing the laser power density (P$_D$), respectively. Cycle III corresponds to the spectra collected at multiple points on the sample, one point to each P$_D$ value—inserts showing the CIE diagram of the obtained emission profile.

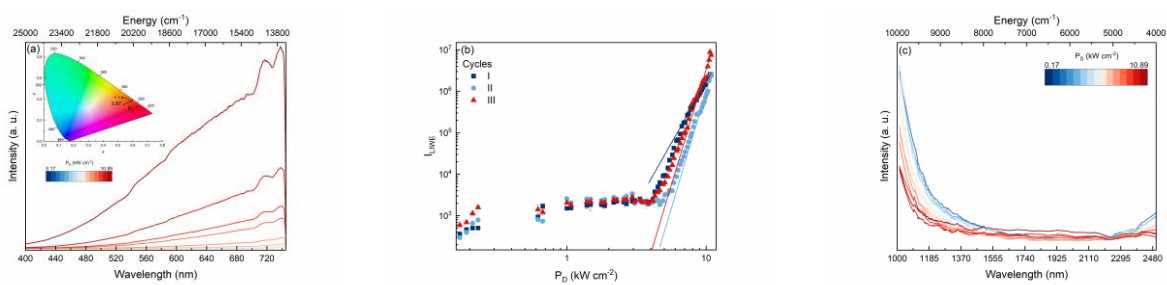

Fig. S6. (a) Photoluminescence spectra of NdMnO$_3$ bulk at $\lambda_{ex}$=808 nm, 10$^3$ mbar and ambient temperature, in function of laser power density (P$_D$) in (a) visible and (c) IR regions. In the IR spectra, note the attenuation effect of the FEL1100 (Thorlabs). (b) Log–log plot of the power density (P$_D$, kW cm$^{-2}$) dependence of the laser-induced white light emission (I$_{LIWE}$) under different cycles of measurements. Cycles I and II correspond to the UC-emission spectra at a single sample spot under increasing and decreasing the laser power density (P$_D$), respectively. Cycle III corresponds to the spectra collected at multiple points on the sample, one point to each P$_D$ value—inserts showing the CIE diagram of the obtained emission profile.

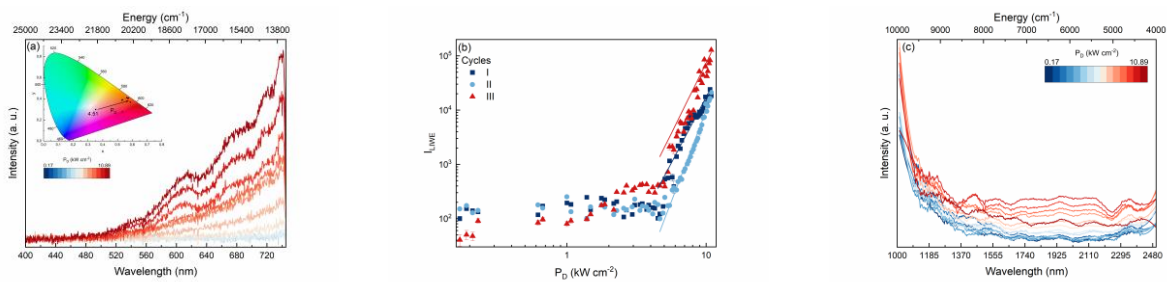

Fig. S7. (a) Photoluminescence spectra of NdMnO$_3$ bulk at $\lambda_{ex}$=808 nm, 10$^{-6}$ mbar and ambient temperature, in function of laser power density (P$_D$) in (a) visible and (c) IR regions. In the IR spectra, note the attenuation effect of the FEL1100 (Thorlabs). (b) Log–log plot of the power density (P$_D$, kW cm$^{-2}$) dependence of the laser-induced white light emission (I$_{LIWE}$) under different cycles of measurements. Cycles I and II correspond to the UC-emission spectra at a single sample spot under increasing and decreasing the laser power density (P$_D$), respectively. Cycle III corresponds to the spectra collected at multiple points on the sample, one point to each P$_D$ value—inserts showing the CIE diagram of the obtained emission profile.

## 5. Anti-Stokes laser-induced white emission (LIWE) at laser excitation in 975 nm: sample compactation and pressure dependencies

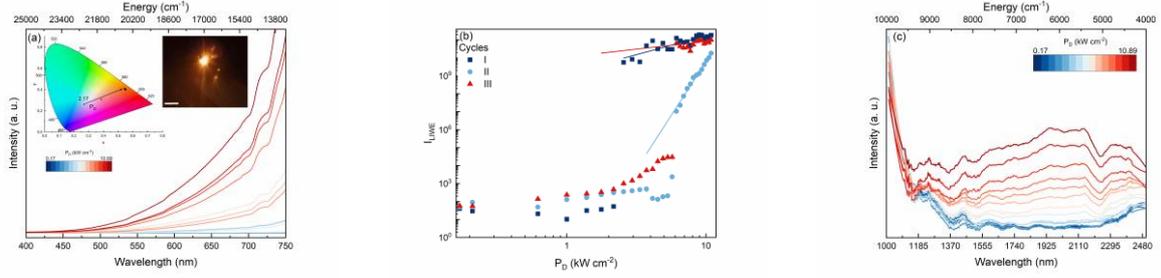

Fig. S8. (a) Photoluminescence spectra of NdMnO₃ nanocrystals at $\lambda_{ex}$=975 nm, $10^3$ mbar and ambient temperature, in function of laser power density ($P_D$) in (a) visible and (c) IR regions. In the IR spectra, note the attenuation effect of the FEL1100 (Thorlabs). (b) Log–log plot of the power density ($P_D$, kW cm⁻²) dependence of the laser-induced white light emission ($I_{LIWE}$) under different cycles of measurements. Cycles I and II correspond to the UC-emission spectra at a single sample spot under increasing and decreasing the laser power density ($P_D$), respectively. Cycle III corresponds to the spectra collected at multiple points on the sample, one point to each $P_D$ value—inserts showing the CIE diagram of the obtained emission profile and photos recorded under 10.89 kW cm⁻².

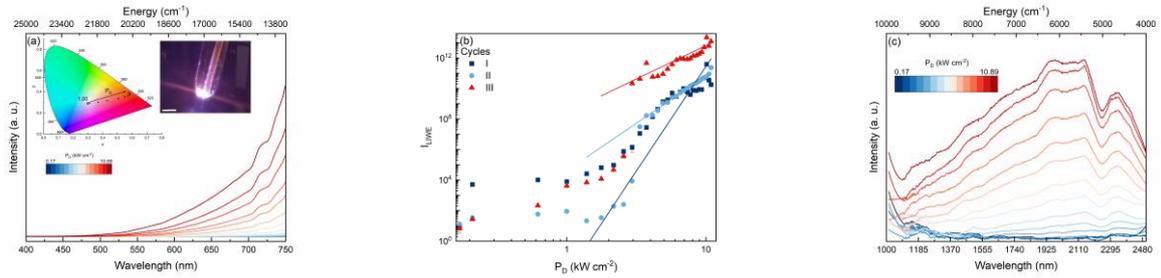

Fig. S9. (a) Photoluminescence spectra for NdMnO₃ nanocrystals at $\lambda_{ex}$=975 nm, $10^{-6}$ mbar and ambient temperature, in function of laser power density ($P_D$) in (a) visible and (c) IR regions. In the IR spectra, note the attenuation effect of the FEL1100 (Thorlabs). (b) Log–log plot of the power density ($P_D$, kW cm⁻²) dependence of the laser-induced white light emission ($I_{LIWE}$) under different cycles of measurements. Cycles I and II correspond to the UC-emission spectra at a single sample spot under increasing and decreasing the laser power density ($P_D$), respectively. Cycle III corresponds to the spectra collected at multiple points on the sample, one point to each $P_D$ value—inserts showing the CIE diagram of the obtained emission profile and photos recorded under 10.89 kW cm⁻².

Tab. S2. Minimum value of laser power density to promote the white light emission ($P_{D, threshold}$, kW cm⁻²), and the slopes of the dependence of the integrated intensity of white light emission ($I_{WL}$) with power density ($P_D$) in a log-log plot of NdMnO₃ analysed in the non-compressed form (nanocrystals) under several excitation conditions $\lambda_{laser\ excitation}$ 808 nm or 975 nm, pressure (P, mbar), temperature (T, K), and different cycles of measurements (cycle I involved measuring the UC emission spectra at a single spot of the sample while increasing $P_D$, whereas cycle II involved measuring the UC emission spectra decreasing $P_D$ at the same place. Cycle III involved measurements at multiple points on the sample. The slope is indicated by n, and the correlation coefficient is $R^2$ in the respective cycle of measurement acquired.

| $\lambda_{laser}$ (nm) | P (mbar) | $P_{D,\ threshold}$ (kW cm⁻²) | n | $R^2$ | Cycle |
|---|---|---|---|---|---|
| **808** | $10^3$ | 4.91 | 1.5 ± 0.3 | 0.6 | I |
| | | 4.91 | 1.4 ± 0.3 | 0.6 | II |
| | | 1.39 | 1.9 ± 0.2 | 0.8 | III |
| | $10^{-6}$ | 2.95 | 1.2 ± 0.5 | 0.5 | I |
| | | 7.34 | 6.7 ± 0.5 | 0.9 | II |
| | | 2.17 | 1.9 ± 0.2 | 0.8 | III |
| **975** | $10^3$ | 2.17 | 2.9 ± 0.3 | 0.7 | I |
| | | 5.31 | 12.0 ± 0.4 | 0.9 | II |
| | | 2.17 | 0.9 ± 0.3 | 0.5 | III |
| | $10^{-6}$ | 1.00 | 13 ± 1 | 0.7 | I |
| | | 1.39 | 6.2 ± 0.2 | 0.9 | II |
| | | 1.77 | 4.3 ± 0.6 | 0.7 | III |

Tab. S3. Minimum value of laser power density to promote the white light emission ($P_{D, \text{threshold}}$, kW cm$^{-2}$), and the slopes of the dependence of the integrated intensity of white light emission ($I_{WL}$) with power density ($P_D$) in a log-log plot of NdMnO$_3$ bulk under several excitation conditions $\lambda_{\text{excitation}}$ 808 nm or 975 nm, pressure (P, mbar), temperature (T, K), and different cycles of measurements (cycle I involved measuring the UC emission spectra at a single spot of the sample while increasing $P_D$, whereas cycle II involved measuring the UC emission spectra decreasing $P_D$ at the same place. Cycle III involved measurements at multiple points on the sample. The slope is indicated by n, and the correlation coefficient is R$^2$ in the respective cycle of measurement acquired.

| $\lambda_{\text{excitation}}$ (nm) | P (mbar) | $P_{D, \text{threshold}}$ (kW cm$^{-2}$) | n | R$^2$ | Cycle |
|---|---|---|---|---|---|
| **808** | 10$^3$ | 3.87 | 5.9 ± 0.1 | 0.9 | I |
| | | 4.65 | 11 ± 0.4 | 0.9 | II |
| | | 3.74 | 11.1 ± 0.7 | 0.9 | III |
| | 10$^{-6}$ | 4.51 | 4.8 ± 0.2 | 0.9 | I |
| | | 4.65 | 6.7 ± 0.2 | 0.9 | II |
| | | 4.65 | 5.2 ± 0.6 | 0.7 | III |
| **975** | 10$^3$ | 2.28 | 3.8 ± 0.1 | 0.9 | I |
| | | 3.87 | 7.2 ± 0,1 | 0.9 | II |
| | | 1.48 | 3,8 ± 0,2 | 0.9 | III |
| | 10$^{-6}$ | 4.65 | 3,9 ± 0,5 | 0.6 | I |
| | | 6.69 | 21 ± 2 | 0.8 | II |
| | | 3.07 | 4.4 ± 0.4 | 0.7 | III |

## 6. Temperature determination for the LIWE

It is noteworthy that the spectral cut-off effect imposed by the filter FESH750 (see Fig. S3) limits the signal acquisition up to 750 nm. This constraint results in spectra that exhibit reduced quality of the fit with the Planck curve distribution and, consequently, lead to the estimation of lower temperatures, as shown in Fig. S10 and Fig. S11.

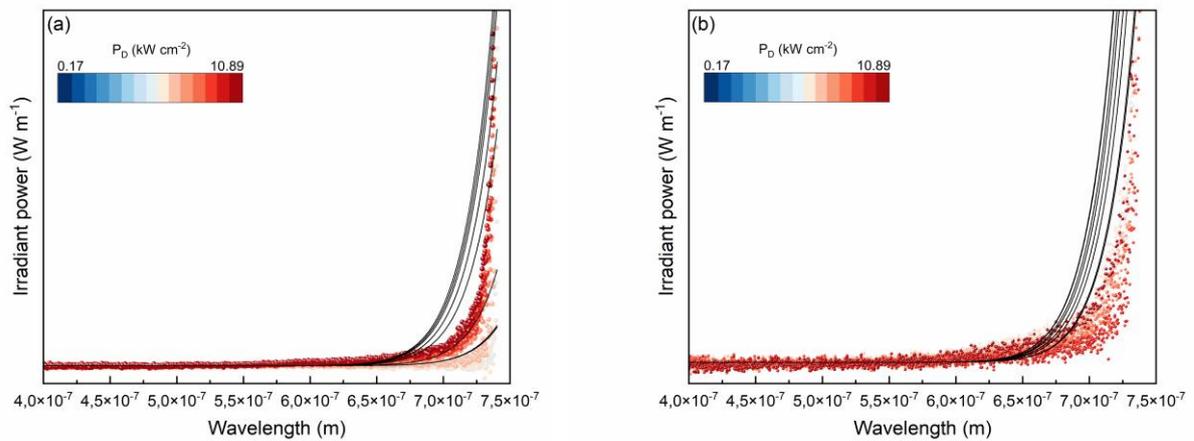

Fig. S10. Fitting of the emission spectra acquired at $\lambda_{\text{ex}}$=808 nm at different values of laser power density ($P_D$) for NdMnO$_3$ analysed in the non-compressed form (nanocrystals) at (a) 10$^3$ mbar, and (b) 10$^{-6}$ mbar, using Planck's law.

Table S4. Fitting results of the continuous upconversion emission spectra for NdMnO$_3$ nanocrystals to Eq. (2) with A = 3.44×10$^{-4}$ and T = 0.014388/B for NdMnO$_3$ (nanocrystals) under 808 nm excitation at ambient pressure (10$^3$ mbar). The fitted temperatures are represented in Fig. S9.

| $P_D$ (kW cm$^{-2}$) | B (×10−5) | B (×10−8) | R$^2$ | T (K) | ΔT (K) |
|---|---|---|---|---|---|
| **4.91** | 2.88 | 7.36 | 0.145 | 500 | 1 |
| **5.05** | 2.82 | 4.08 | 0.351 | 511.1 | 0.7 |
| **5.31** | 2.79 | 2.81 | 0.530 | 516.6 | 0.5 |
| **5.46** | 2.77 | 2.67 | 0.550 | 519.5 | 0.5 |
| **5.72** | 2.77 | 2.46 | 0.581 | 518.8 | 0.5 |
| **5.87** | 2.73 | 2.78 | 0.523 | 527.5 | 0.5 |

| 6.14 | 2.76 | 3.63 | 0.407 | 520.5 | 0.7 |
|---|---|---|---|---|---|
| 6.29 | 2.74 | 2.12 | 0.660 | 525.4 | 0.4 |
| 6.53 | 2.73 | 2.23 | 0.634 | 527.5 | 0.4 |
| 6.69 | 2.72 | 1.78 | 0.727 | 529.0 | 0.3 |
| 6.91 | 2.71 | 1.75 | 0.733 | 531.5 | 0.3 |
| 7.07 | 2.70 | 1.47 | 0.794 | 533.8 | 0.3 |
| 7.34 | 2.68 | 1.90 | 0.700 | 536.8 | 0.4 |
| 7.50 | 2.71 | 2.77 | 0.505 | 530.5 | 0.5 |
| 7.73 | 2.64 | 1.57 | 0.770 | 544.5 | 0.3 |
| 7.89 | 2.65 | 1.19 | 0.851 | 543.5 | 0.2 |
| 8.11 | 2.66 | 1.25 | 0.833 | 541.7 | 0.3 |
| 8.27 | 2.67 | 1.67 | 0.740 | 538.7 | 0.3 |
| 8.55 | 2.65 | 1.25 | 0.834 | 543.2 | 0.3 |
| 8.71 | 2.67 | 1.80 | 0.710 | 538.2 | 0.4 |
| 8.94 | 2.66 | 1.47 | 0.784 | 540.9 | 0.3 |
| 9.11 | 2.67 | 2.90 | 0.477 | 538.2 | 0.6 |
| 9.34 | 2.66 | 1.29 | 0.826 | 540.4 | 0.3 |
| 9.51 | 2.66 | 1.37 | 0.806 | 541.3 | 0.3 |
| 9.72 | 2.66 | 2.24 | 0.605 | 540.7 | 0.5 |
| 9.89 | 2.65 | 1.27 | 0.827 | 543.4 | 0.3 |
| 10.12 | 2.62 | 1.22 | 0.847 | 549.8 | 0.3 |
| 10.29 | 2.65 | 1.78 | 0.710 | 542.4 | 0.4 |
| 10.51 | 2.65 | 1.52 | 0.771 | 543.4 | 0.3 |
| 10.68 | 2.65 | 1.50 | 0.775 | 543.4 | 0.3 |
| 10.89 | 2.59 | 2.04 | 0.668 | 556.2 | 0.4 |

Table S5. Fitting results of the continuous upconversion emission spectra for NdMnO$_3$ nanocrystals to Eq. (2) with A = 3.24×10$^{-12}$ and T = 0.014388/B under 808 nm excitation at low pressure (10$^{-6}$ mbar). The fitted temperatures are represented in Fig. S9.

| P$_D$ (kW cm$^{-2}$) | B (×10−5) | B (×10−8) | R$^2$ | T (K) | ΔT (K) |
|---|---|---|---|---|---|
| 3.74 | 2.87 | 3.67 | 0.396 | 501.9 | 0.6 |
| 3.87 | 2.86 | 3.75 | 0.386 | 503.1 | 0.7 |
| 4.14 | 2.84 | 4.53 | 0.301 | 506.7 | 0.8 |
| 4.28 | 2.83 | 4.02 | 0.351 | 509.1 | 0.7 |
| 4.51 | 2.82 | 4.13 | 0.338 | 510.8 | 0.7 |
| 4.65 | 2.81 | 3.46 | 0.421 | 511.3 | 0.6 |
| 4.91 | 2.81 | 3.70 | 0.389 | 511.7 | 0.7 |
| 5.05 | 2.80 | 4.20 | 0.331 | 513.5 | 0.8 |
| 5.31 | 2.80 | 3.81 | 0.370 | 514.3 | 0.7 |
| 5.46 | 2.80 | 3.82 | 0.372 | 514.5 | 0.7 |
| 5.72 | 2.79 | 3.89 | 0.365 | 515.2 | 0.7 |
| 5.87 | 2.78 | 3.62 | 0.397 | 517.1 | 0.7 |
| 6.14 | 2.73 | 2.63 | 0.410 | 527.0 | 0.9 |
| 6.29 | 2.78 | 3.52 | 0.376 | 517.0 | 0.7 |
| 6.53 | 2.78 | 3.85 | 0.368 | 517.3 | 0.7 |
| 6.69 | 2.79 | 3.54 | 0.408 | 515.8 | 0.7 |
| 6.91 | 2.79 | 3.97 | 0.354 | 515.3 | 0.7 |
| 7.07 | 2.80 | 4.29 | 0.322 | 513.2 | 0.8 |
| 7.34 | 2.81 | 3.82 | 0.375 | 512.9 | 0.7 |
| 7.50 | 2.81 | 3,89 | 0.365 | 511.8 | 0,7 |
| 7.73 | 2.81 | 3.99 | 0.355 | 512.5 | 0.7 |
| 7.89 | 2.81 | 3.75 | 0.384 | 512.1 | 0.7 |
| 8.11 | 2.81 | 3.72 | 0.386 | 511.4 | 0.7 |
| 8.27 | 2.80 | 3.69 | 0.389 | 513.1 | 0.7 |
| 8.55 | 2.81 | 3.81 | 0.375 | 511.5 | 0.7 |
| 8.71 | 2.82 | 3.65 | 0.396 | 510.0 | 0.7 |
| 8.94 | 2.84 | 3.65 | 0.397 | 507.2 | 0.7 |
| 9.11 | 2.84 | 3.69 | 0.392 | 506.4 | 0.7 |
| 9.34 | 2.84 | 4.40 | 0.313 | 506.8 | 0.8 |
| 9.51 | 2.84 | 4.11 | 0.343 | 507.1 | 0.7 |
| 9.72 | 2.84 | 3.93 | 0.362 | 506.5 | 0.7 |
| **9.89** | 3.02 | 7.42 | 0.147 | 476.0 | 1.2 |
| **10.12** | 2.84 | 4.33 | 0.320 | 507.3 | 0.8 |
| **10.29** | 2.82 | 4.06 | 0.347 | 509.5 | 0.7 |
| **10.51** | 2.81 | 3.79 | 0.377 | 512.6 | 0.7 |
| **10.68** | 2.80 | 3.63 | 0.397 | 514.6 | 0.7 |
| **10.89** | 2.78 | 3.96 | 0.36 | 516.9 | 0.7 |

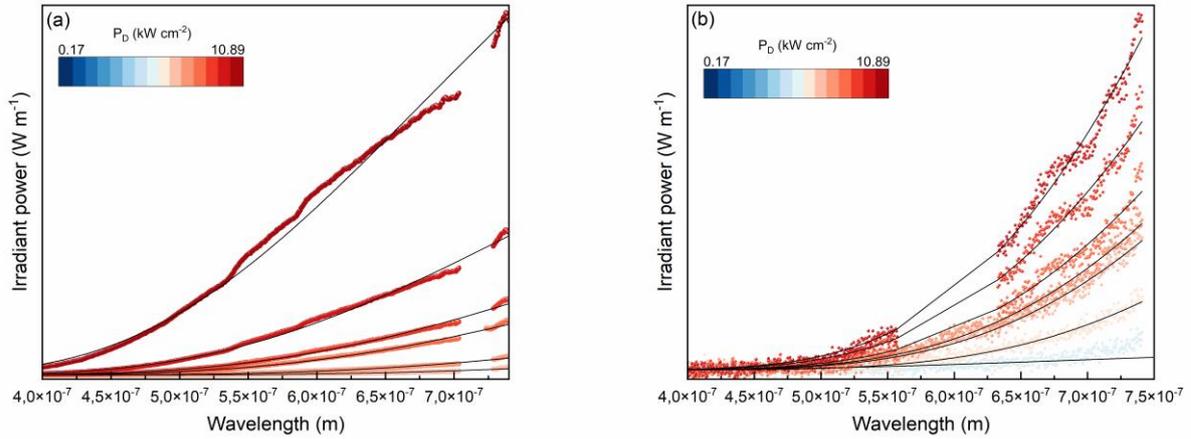

Fig. S11 Fitting of the emission spectra acquired at $\lambda_{ex}$=808 nm at different values of laser power density ($P_D$) for NdMnO$_3$ bulk at (a) 10$^3$ mbar, and (b) 10$^{-6}$ mbar, using Planck's law.

Table S6. Fitting results of the continuous upconversion emission spectra for NdMnO$_3$ bulk to Eq. (2) with A = 9.00×10$^{-24}$ and T = 0.014388/B for NdMnO$_3$ (ceramic) under 808 nm excitation at ambient pressure (10$^3$ mbar). The fitted temperatures are represented in Fig. S10.

| $P_D$ (kW cm$^{-2}$) | B (×10−6) | B (×10−9) | R$^2$ | T (K) | ΔT (K) |
|---|---|---|---|---|---|
| **6.91** | 8.71 | 3.82 | 0.971 | 1652.7 | 0.7 |
| **7.07** | 8.53 | 3.07 | 0.980 | 1687.5 | 0.6 |
| **7.34** | 8.36 | 2.59 | 0.986 | 1721.6 | 0.5 |
| **7.50** | 8.17 | 2.05 | 0.991 | 1762.0 | 0.4 |
| **7.73** | 7.98 | 1.80 | 0.993 | 1802.7 | 0.4 |
| **7.89** | 7.85 | 1.63 | 0.994 | 1832.6 | 0.4 |
| **8.11** | 7.62 | 1.34 | 0.996 | 1889.1 | 0.3 |
| **8.27** | 7.42 | 1.19 | 0.996 | 1938.3 | 0.3 |
| **8.55** | 7.17 | 1.13 | 0.996 | 2007.5 | 0.3 |
| **8.71** | 7.17 | 1.13 | 0.997 | 2007.5 | 0.3 |
| **8.94** | 7.12 | 1.09 | 0.997 | 2050.5 | 0.3 |
| **9.11** | 7.06 | 1.04 | 0.997 | 2095.5 | 0.3 |
| **9.34** | 6.91 | 1.08 | 0.997 | 2140.6 | 0.3 |
| **9.51** | 6.79 | 1.10 | 0.997 | 2185.6 | 0.3 |
| **9.72** | 6.70 | 1.03 | 0.997 | 2230.6 | 0.3 |
| **9.89** | 6.42 | 1.08 | 0.997 | 2275.6 | 0.4 |
| **10.12** | 6.42 | 1.08 | 0.997 | 2320.7 | 0.4 |
| **10.29** | 6.33 | 1.05 | 0.997 | 2365.7 | 0.4 |
| **10.51** | 5.96 | 1.30 | 0.994 | 2414.8 | 0.5 |
| **10.68** | 5.83 | 1.35 | 0.994 | 2469.5 | 0.6 |
| **10.89** | 5.72 | 1.33 | 0.994 | 2514.7 | 0.6 |

Table S7. Fitting results of the continuous upconversion emission spectra for NdMnO$_3$ analysed in the bulk to Eq. (2) with A = 1.15×10$^{-24}$ and T = 0.014388/B, under 808 nm excitation at low pressure (10$^{-6}$ mbar). The fitted temperatures are represented in Fig. S10.

| $P_D$ (kW cm$^{-2}$) | B (×10−6) | B (×10−8) | R$^2$ | T (K) | ΔT (K) |
|---|---|---|---|---|---|
| **6.14** | 9.46 | 1.30 | 0.735 | 1520.3 | 2 |
| **6.29** | 9.26 | 1.03 | 0.800 | 1553.0 | 2 |
| **6.53** | 9.05 | 0.71 | 0.903 | 1589.4 | 1 |
| **6.69** | 8.80 | 0.54 | 0.940 | 1634.6 | 1 |
| **6.91** | 8.77 | 0.49 | 0.943 | 1640.6 | 1 |
| **7.07** | 8.08 | 0.50 | 0.975 | 1781.1 | 1 |
| **7.34** | 8.40 | 0.40 | 0.971 | 1713.3 | 0.8 |
| **7.50** | 8.39 | 0.30 | 0.978 | 1714.7 | 0.6 |
| **7.73** | 8.30 | 0.29 | 0.980 | 1734.1 | 0.6 |
| **7.89** | 8.27 | 0,27 | 0.983 | 1740.2 | 0.6 |
| **8.11** | 8.21 | 0.25 | 0.984 | 1753.5 | 0.5 |
| **8.27** | 8.15 | 0.24 | 0.986 | 1766.4 | 0.5 |
| **8.55** | 8.21 | 0.25 | 0.985 | 1753.5 | 0.5 |
| **8.71** | 8.15 | 0.24 | 0.986 | 1766.0 | 0.5 |
| **8.94** | 8.08 | 0.23 | 0.990 | 1781.7 | 0.5 |
| **9.11** | 8.12 | 0.25 | 0.988 | 1771.6 | 0.6 |
| **9.34** | 8.06 | 0.26 | 0.987 | 1785.5 | 0.6 |
| **9.51** | 7.99 | 0.25 | 0.989 | 1800.7 | 0.6 |
| **9.72** | 7.91 | 0.24 | 0.989 | 1818.1 | 0.6 |
| **9.89** | 7.68 | 0.18 | 0.994 | 1872.2 | 0.4 |
| **10.12** | 7.81 | 0.21 | 0.992 | 1841.4 | 0.5 |
| **10.29** | 7,70 | 0.20 | 0.993 | 1869.0 | 0.5 |
| **10.51** | 7.67 | 0.19 | 0.993 | 1877.1 | 0.5 |
| **10.68** | 7.47 | 0.15 | 0.996 | 1927.0 | 0.4 |
| **10.89** | 6.93 | 0.08 | 0.495 | 2075.6 | 0.2 |

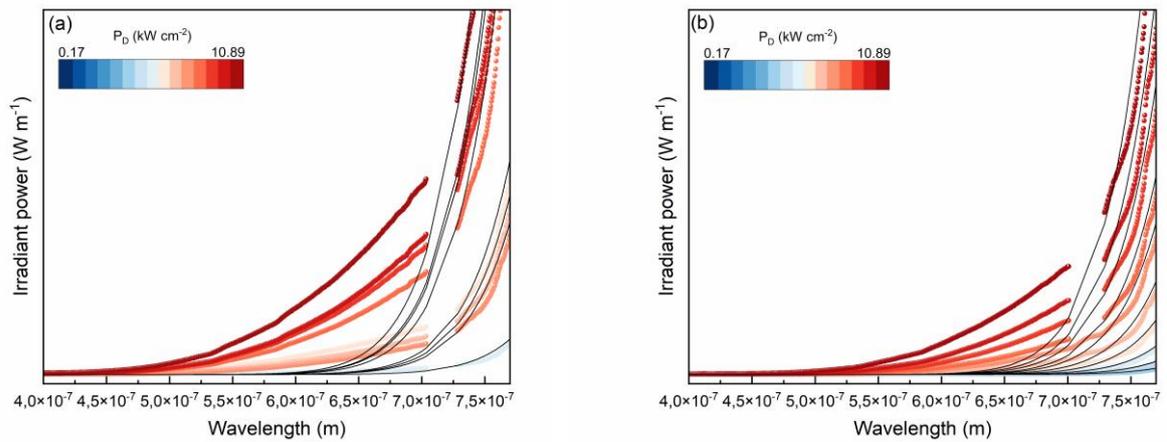

Fig. S12 Fitting of the emission spectra acquired at $\lambda_{ex}$=975 nm at different values of laser power density ($P_D$) for NdMnO$_3$ nanocrystals at (a) $10^3$ mbar, and (b) $10^{-6}$ mbar, using Planck's law.

Table S8. Fitting results of the continuous upconversion emission spectra for NdMnO$_3$ nanocrystals to Eq. (2) with A = $8.00\times10^{-17}$ and T = 0.014388/B for NdMnO$_3$ nanocrystals under 975 nm excitation at ambient pressure ($10^3$ mbar). The fitted temperatures are represented in Fig. S11.

| P$_D$ (kW cm$^{-2}$) | B (×10−5) | B (×10−8) | R$^2$ | T (K) | ΔT (K) |
|---|---|---|---|---|---|
| **5.31** | 2.21 | 1.03 | 0.873 | 650.2 | 0.3 |
| **5.72** | 2.14 | 0.97 | 0.887 | 671.3 | 0.3 |
| **6.14** | 2.07 | 1.00 | 0.870 | 693.7 | 0.3 |
| **6.53** | 2.09 | 0.96 | 0.880 | 689.0 | 0.3 |
| **6.91** | 2.09 | 0.93 | 0.888 | 689.1 | 0.3 |
| **7.34** | 2.08 | 0.93 | 0.890 | 692.8 | 0.3 |
| **7.73** | 2.10 | 0.90 | 0.896 | 685.0 | 0.3 |
| **8.11** | 2.04 | 0.91 | 0.895 | 704.9 | 0.3 |
| **8.55** | 2.01 | 0.91 | 0.894 | 716.5 | 0.3 |
| **8.94** | 1.99 | 0.90 | 0.895 | 721.3 | 0.3 |
| **9.34** | 1.99 | 0.89 | 0.897 | 722.6 | 0.3 |

| $P_D$ (kW cm$^{-2}$) | B (×10−5) | B (×10−8) | $R^2$ | T (K) | ΔT (K) |
|---|---|---|---|---|---|
| 9.72 | 2.00 | 0.85 | 0.905 | 719.1 | 0.3 |
| 10.12 | 1.99 | 0.86 | 0.903 | 724.1 | 0.3 |
| 10.51 | 1.98 | 0.85 | 0.904 | 724.9 | 0.3 |
| 10.89 | 1.97 | 0.87 | 0.899 | 731.6 | 0.3 |

Table S9. Fitting results of the continuous upconversion emission spectra for NdMnO$_3$ nanocrystals to Eq. (2) with A = 2.70×10$^{-16}$ and T = 0.014388/B for NdMnO$_3$ (nanocrystals) under 975 nm excitation at low pressure (10$^{-6}$ mbar). The fitted temperatures are represented in Fig. S11.

| $P_D$ (kW cm$^{-2}$) | B (×10−5) | B (×10−8) | $R^2$ | T (K) | ΔT (K) |
|---|---|---|---|---|---|
| 3.74 | 2.43 | 1.02 | 0.871 | 590.3 | 0.2 |
| 4.14 | 2.41 | 0.91 | 0.895 | 597.3 | 0.2 |
| 4.51 | 2.38 | 0.88 | 0.901 | 604.2 | 0.2 |
| 4.91 | 2.34 | 0.78 | 0.920 | 613.3 | 0.2 |
| 5.31 | 2.32 | 0.79 | 0.918 | 619.5 | 0.2 |
| 5.72 | 2.29 | 0.80 | 0.916 | 628.6 | 0.2 |
| 6.14 | 2.27 | 0.77 | 0.922 | 633.1 | 0.2 |
| 6.53 | 2.24 | 0.78 | 0.920 | 639.6 | 0.2 |
| 6.91 | 2.23 | 0.78 | 0.918 | 644.7 | 0.2 |
| 7.34 | 2.21 | 0.78 | 0.917 | 650.5 | 0.2 |
| 7.73 | 2.19 | 0.78 | 0.917 | 655.7 | 0.2 |
| 8.11 | 2.17 | 0.78 | 0.917 | 660.0 | 0.2 |
| 8.55 | 2.17 | 0.82 | 0.919 | 662.0 | 0.2 |
| 8.94 | 2.16 | 0.80 | 0.913 | 665.8 | 0.2 |
| 9.34 | 2.14 | 0.83 | 0.915 | 671.1 | 0.3 |
| 9.72 | 2.13 | 0.82 | 0.908 | 674.8 | 0.3 |
| 10.12 | 2.12 | 0.85 | 0.911 | 677.8 | 0.3 |
| 10.51 | 2.11 | 0.85 | 0.901 | 681.9 | 0.3 |
| 10.89 | 2.09 | 0.89 | 0.899 | 685.7 | 0.3 |

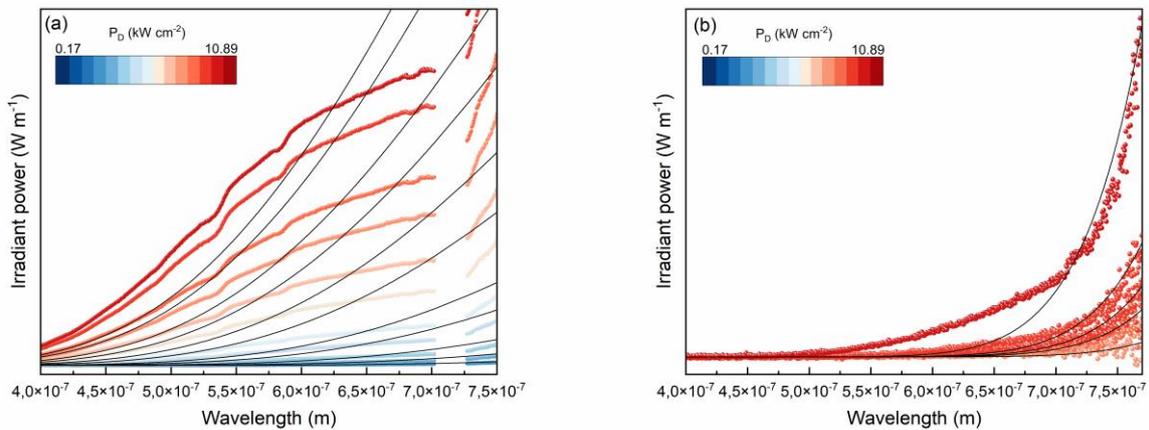

Fig. S13. Fitting of the emission spectra acquired at λ$_{ex}$=975 nm at different values of laser power density (P$_D$) for NdMnO$_3$ analysed in the compressed form (ceramic) at (a) 10$^3$ mbar, and (b) 10$^{-6}$ mbar, using Planck's law.

Table S10. Fitting results of the continuous upconversion emission spectra for NdMnO$_3$ bulk to Eq. (2) with A = 6.65×10$^{-16}$ and T = 0.014388/B, under 975 nm excitation at ambient pressure (10$^3$ mbar). The fitted temperatures are represented in Fig. X in the manuscript.

| $P_D$ (kW cm$^{-2}$) | B (×10−6) | B (×10−9) | $R^2$ | T (K) | ΔT (K) |
|---|---|---|---|---|---|
| 3.74 | 8.19 | 9.16 | 0.859 | 1756 | 2 |
| 3.87 | 8.08 | 9.03 | 0.862 | 1781 | 2 |
| 4.14 | 7.86 | 9.16 | 0.851 | 1831 | 2 |
| 4.28 | 7.82 | 9.86 | 0.834 | 1840 | 2 |
| 4.51 | 7.59 | 9.32 | 0.842 | 1895 | 2 |
| 4.65 | 7.50 | 9.55 | 0.834 | 1917 | 2 |
| 4.91 | 7.41 | 9.38 | 0.836 | 1941 | 2 |
| 5.05 | 7.17 | 9.49 | 0.826 | 2007 | 3 |
| 5.31 | 7.02 | 9.60 | 0.818 | 2050 | 3 |
| 5.46 | 6.83 | 9.45 | 0.813 | 2105 | 3 |
| 5.72 | 6.73 | 9.27 | 0.814 | 2139 | 3 |
| 5.87 | 6.47 | 9.43 | 0.803 | 2223 | 3 |
| 6.14 | 6.72 | 9.71 | 0.766 | 2142 | 3 |
| 6.29 | 6.57 | 9.39 | 0.807 | 2191 | 3 |
| 6.53 | 6.52 | 9.42 | 0.805 | 2205 | 3 |
| 6.69 | 6.41 | 9.50 | 0.799 | 2243 | 3 |
| 6.91 | 6.26 | 9.41 | 0.796 | 2299 | 3 |
| 7.07 | 6.17 | 9.30 | 0.797 | 2332 | 4 |
| 7.34 | 6.11 | 9.29 | 0.795 | 2354 | 4 |
| 7.50 | 6.09 | 9.29 | 0.795 | 2361 | 4 |
| 7.73 | 6.01 | 9.33 | 0.790 | 2394 | 4 |
| 7.89 | 5.97 | 9.34 | 0.789 | 2412 | 4 |
| 8.11 | 5.96 | 9.31 | 0.789 | 2414 | 4 |
| 8.27 | 5.84 | 9.33 | 0.787 | 2466 | 4 |
| 8.55 | 5.76 | 9.21 | 0.790 | 2499 | 4 |
| 8.71 | 5.71 | 9.28 | 0.788 | 2518 | 4 |
| 8.94 | 5.80 | 9.26 | 0.791 | 2482 | 4 |
| 9.11 | 5.69 | 9.18 | 0.791 | 2527 | 4 |
| 9.34 | 5.60 | 9.29 | 0.785 | 2569 | 4 |
| 9.51 | 5.58 | 9.23 | 0.784 | 2576 | 4 |
| 9.72 | 5.59 | 9.23 | 0.785 | 2572 | 4 |
| 9.89 | 5.48 | 9.17 | 0.782 | 2624 | 4 |
| 10.12 | 5.38 | 9.11 | 0.782 | 2676 | 5 |
| 10.29 | 5.49 | 9.01 | 0.765 | 2620 | 4 |
| 10.51 | 5.54 | 9.23 | 0.783 | 2599 | 4 |
| 10.68 | 5.43 | 9.20 | 0.781 | 2650 | 4 |
| 10.89 | 5.29 | 9.10 | 0.779 | 2722 | 5 |

Table S11. Fitting results of the continuous upconversion emission spectra for $NdMnO_3$ bulk to Eq. (2) with $A = 1.20 \times 10^{-12}$ and $T = 0.014388/B$, under 975 nm excitation at low pressure ($10^{-6}$ mbar). The fitted temperatures are represented in Fig. X in the manuscript.

| $P_D$ (kW cm$^{-2}$) | B (×10−6) | B (×10−8) | $R^2$ | T (K) | ΔT (K) |
|---|---|---|---|---|---|
| 7.73 | 1.75 | 5.32 | 0.147 | 824 | 3 |
| 7.89 | 1.71 | 2.75 | 0.458 | 843 | 1 |
| 8.11 | 1.66 | 1.95 | 0.632 | 868 | 1 |
| 8.27 | 1.67 | 1.39 | 0.763 | 862.8 | 0.7 |
| 8.55 | 1.66 | 1.33 | 0.779 | 868.7 | 0.7 |
| 8.71 | 1.63 | 1.31 | 0.784 | 880.6 | 0.7 |
| 8.94 | 1.63 | 1.28 | 0.796 | 883.1 | 0.7 |
| 9.11 | 1.60 | 1.07 | 0.842 | 897.5 | 0.6 |
| 9.34 | 1.63 | 1.02 | 0.858 | 885.2 | 0.6 |
| 9.51 | 1.58 | 0.95 | 0.873 | 908.0 | 0.5 |
| 9.72 | 1.59 | 0.92 | 0.881 | 904.0 | 0.5 |
| 9.89 | 1.55 | 0.76 | 0.913 | 925.5 | 0.5 |
| 10.12 | 1.59 | 0.82 | 0.900 | 905.8 | 0.5 |
| 10.29 | 1.57 | 0.95 | 0.873 | 916.5 | 0.6 |
| 10.51 | 1.57 | 0.85 | 0.896 | 918.0 | 0.5 |
| 10.68 | 1.54 | 0.88 | 0.886 | 932.0 | 0.5 |
| 10.89 | 1.51 | 0.77 | 0.907 | 955.3 | 0.5 |

Table S12. Slopes of the dependence of the calculated temperature, Eq. (2), with $P_D$ of NdMnO$_3$ in nanocrystals and bulk at different pressures, are represented in Fig.4 in the manuscript.

| Sample | $\lambda_{ex}$ | Slope | $R^2$ | P (mbar) |
|--------|------|-------|-------|----------|
| nanocrystals | 808 | 4.9 ± 0.5 | 0.72 | $10^3$ |
| | | 0.02 ± 0.3 | 0.02 | $10^{-6}$ |
| | 975 | 14.2 ± 0.6 | 0.95 | $10^3$ |
| | | 13.1 ± 0.3 | 0.98 | $10^{-6}$ |
| ceramic | 808 | 222 ± 4 | 0.99 | $10^3$ |
| | | 84 ± 5 | 0.89 | $10^{-6}$ |
| | 975 | 140 ± 3 | 0.98 | $10^3$ |
| | | 29 ± 3 | 0.86 | $10^{-6}$ |

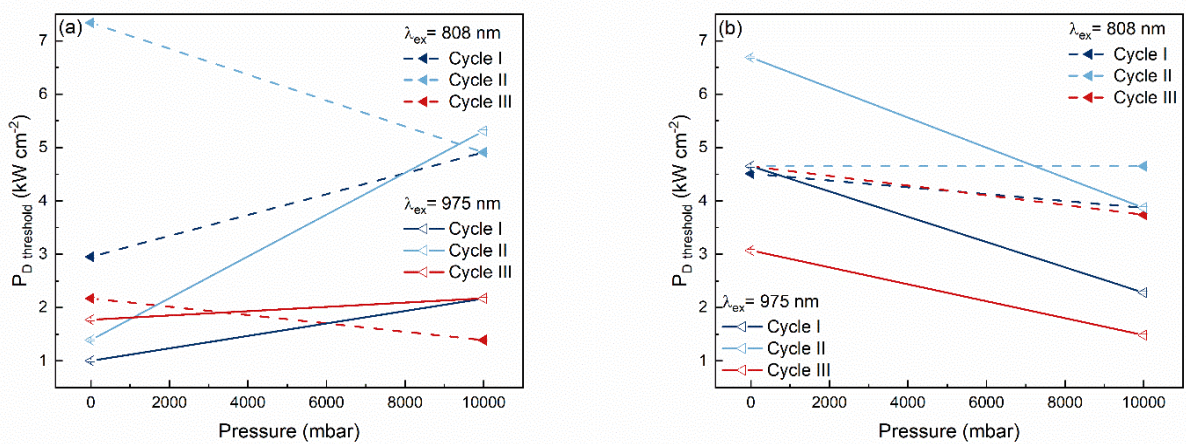

Fig. S14. Dynamics of the threshold due to the different cycles of excitation to NdMnO$_3$ in (a) nanocrystals and (b) bulk samples at $\lambda_{ex}$=808 or 975 nm. Cycles I and II correspond to the UC-emission spectra at a single sample spot under increasing and decreasing the laser power density ($P_D$), respectively. Cycle III corresponds to the spectra collected at multiple points on the sample, one for each $P_D$ value.

The bulk is a ceramic in a circular shape with a diameter of 0.5 cm and has an area of 0.19 cm². The spot size of the laser beam used in the excitation is 0.0001 cm², as discussed in the SI. Therefore, the temperature calculated from the emission spectra corresponds specifically to the localised region directly under laser irradiation, which represents only 0.07% of the total sample area. As such, it does not reflect the bulk temperature of the sample. For instance, a spectrally determined temperature of 2000 K pertains solely to the confocal centre of the irradiated spot (~$10^{-4}$ cm²) and should not be interpreted as the temperature measured by a thermocouple [16]

## 7. Kinetics of the LIWE

The emission spectra for the temporal analysis were acquired under the identical experimental setups used to study the LIWE, shown in Fig. S3, under different pressure conditions, and based on the methodology described previously.[25,26] We performed the kinetics, under laser excitation at 975 nm, and two conditions of power density ($P_D$), firstly, at the threshold to white emission, at 6.91 kW cm⁻², and the $P_D$ maximum at 10.89 kW cm⁻². Following the data acquisition procedure:

1) Set the laser at fixed power density and wait 30 minutes for stabilisation (shutter closed).

2) Set parameters for spectral acquisition, considering the minimum value of the integration time to the detector response.

3) Start measurement.

4) Open the laser shutter at a fixed time.

5) Close the laser shutter at a fixed time.

6) End of measurement at a fixed time, closing the shutter.

7) Save data and check the transient curve.

8) Optimise parameters if needed.

9) Repeat steps 3-7 (3 times to each measurement cycle).

The error bars on the graphs were determined through the integration time used on the spectra acquisition and the integrated intensity of the laser-induced white light emission.

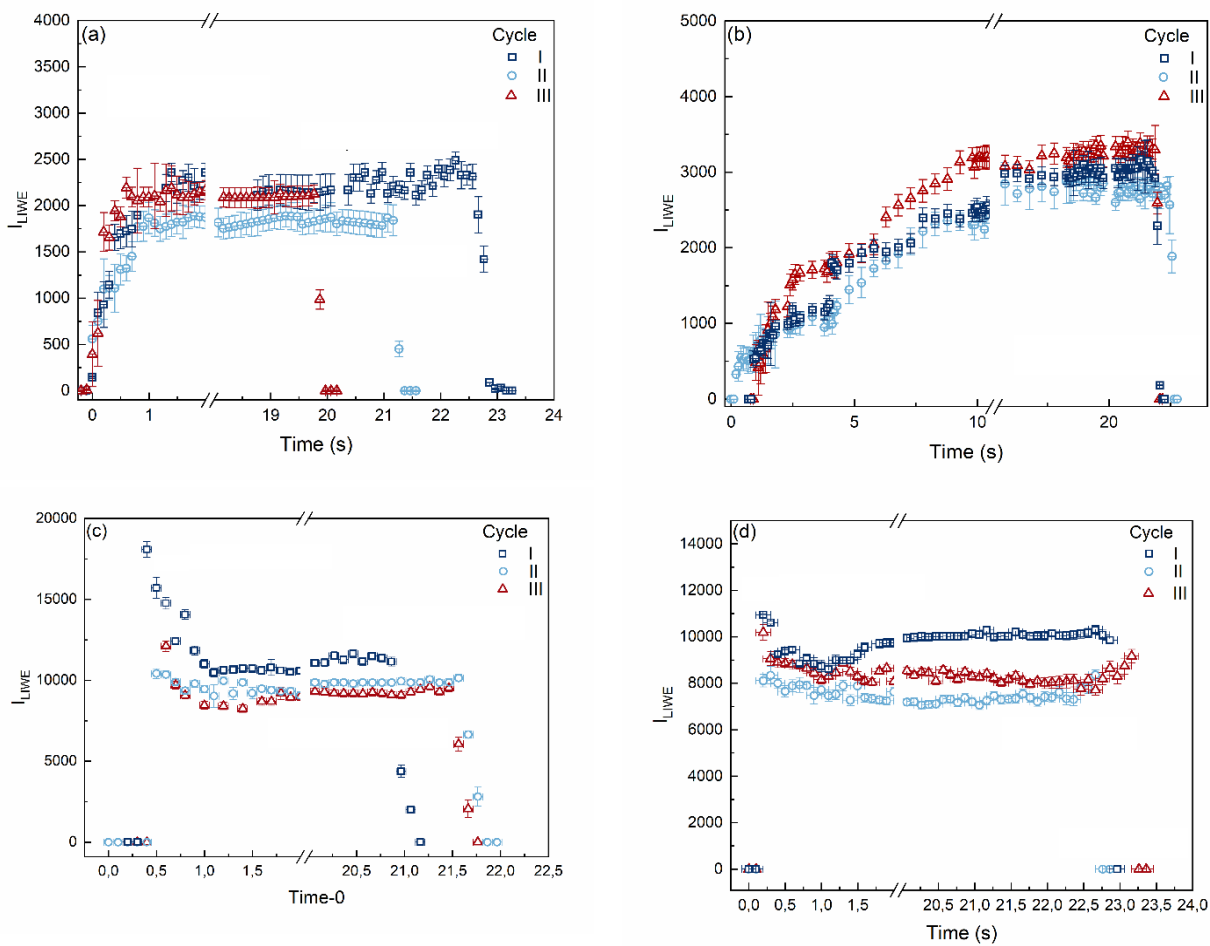

Fig. S15. Transient integrated intensity over 400 at 750 nm at 6.91 kW cm$^{-2}$ (a)10$^3$ mbar, (b) 10$^{-6}$ mbar and at 10.89 kW cm$^{-2}$ (c)10$^3$ mbar, and (d) 10$^{-6}$ mbar.

## 8. Modelling LIWE

Determination of ionisation fractions to Nd$^{3+}$ + h$\nu \rightarrow$ Nd$^{4+}$ + e$^-$, from ionisation energies ~0 until 52.58 eV, excitation at 808 and 975 nm, and different pressure conditions. Firstly, the calculation of the n$_e$ is the electron number density (in cm$^{-3}$) of the Nd$^{3+}$ in the NdMnO$_3$:

Molar Mass of NdMnO$_3$= 247.18 g mol$^{-1}$

Mass Density ($\rho$)= 6.479 Mg m$^{-3}$ (see reference[27])

n$_{NdMnO3}$ = M/$\rho$ corresponds to the formula in units per cm$^3$

Since each NdMnO$_3$ formula unit contains one Nd$^{3+}$ ion, the number density of Nd$^{3+}$ ions is:

n$_{Nd^{3+}}$ = n$_{NdMnO_3}$= (M/ $\rho$) × NA, NA is Avogadro's number (6.022 × 10$^{23}$ mol$^{-1}$)

$n_{Nd^{3+}}$ = (247.18/ 6.479) × 6.022 × $10^{23}$

$n_{Nd^{3+}}$ = 0.02622 × 6.022 × $10^{23}$

$n_{Nd^{3+}}$ =1.578 × $10^{22}$ ions $cm^{-1}$

Each $Nd^{3+}$ ion has three 4f-valence electrons that contribute to the total electron number density ($n_e$):

$n_e$ =3 × $n_{Nd^{3+}}$

Thus,

$n_e$ =3 × $n_{Nd^{3+}}$

$n_e$ =3 × 1.578 × $10^{22}$ electrons $cm^{-3}$

$n_e$ = 4.734 × $10^{22}$ electrons $cm^{-3}$

The ionisation fraction to $Nd^{3+}$ + hν→ $Nd^{4+}$ + $e^-$, as described previously,[28,29] is calculated by Eq. S6, and here, considering the electron number density ($n_e$) 4.734 × $10^{22}$ electrons $cm^{-3}$ generated from the $Nd^{3+}$ into the $NdMnO_3$ solid sample.

$$\frac{N_{i+1}}{N_i} = \frac{2}{n_e}\left(\frac{2\pi m_e k_B T}{h^2}\right)^{3/2} e^{-E_{ion}/k_B T}$$  Eq. S6

$N_{i+1}$ and $N_i$ are the number densities of $Nd^{4+}$ and $Nd^{3+}$, respectively.

$n_e$ is the electron number density (calculated as 4.734 × $10^{22}$ electrons $cm^{-3}$ to $Nd^{3+}$ in the solid sample of $NdMnO_3$),

$m_e$ is the electron mass is (9.109×$10^{-31}$ kg),

$k_B$ is Boltzmann's constant (8.617×$10^{-5}$ eV $K^{-1}$),

h is Planck's constant (6.626×$10^{-34}$ J·s),

T is the temperature (in K) corresponding to the laser power density excitation value,

$E_{ion}$ is the ionisation energy of $Nd^{3+}$ to $Nd^{4+}$, considering Eion ~Eg = 0.46 eV (Keldysh approximation, see reference 29).

Calculating the thermal de Broglie wavelength term:

$\left(\frac{2\pi m_e k_B T}{h^2}\right)^{3/2}$ = 8.0144x$10^{-4}$

And the exponential term:

$e^{-E_{ion}/k_B T} \sim 5.2530 x 10^{-78}$

Then, the ionisation fraction to the of $Nd^{3+}$ + hν→ $Nd^{4+}$ + $e^-$, under excitation at 808 and 975 nm, from the 0.17 to 10.89 kW $cm^{-2}$, at $10^{-6}$ to $10^3$ mbar, are shown in Fig. 5a in the main manuscript to the $E_{ion}$~0.46 eV, and in Fig.S15 to the $E_{ion}$~4.6 eV, as determined previously.[10] The electron density ($n_e$) promoted by $Mn^{2+}$ and $Mn^{4+}$ are 3.156 × $10^{22}$ electrons $cm^{-3}$ and 6.312 × $10^{22}$ electrons $cm^{-3}$, respectively. The ionisation fractions to $Mn^{3+}$ + hν→ $Mn^{4+}$ + $e^-$, and $Mn^{2+}$ + hν→ $Mn^{3+}$ + $e^-$, shown in Fig. b and c, respectively, were calculated considering the ionisation energies from ~0.46 eV, $P_D$ from 0.17 to 10.89 kW$cm^{-2}$, excitation at 808 and 975 nm, and at different pressure conditions. Considering also the ionisation energy of the Mn-ions, reported as >50 eV,[30] the exponential term in the Eq. S4 is so small that it cannot be computed. The ionisation rate values, ranging from $10^{-209}$ to $10^{-31}$, can be considered negligible, indicating that the multiphoton ionisation process does not occur at the origin of LIWE.

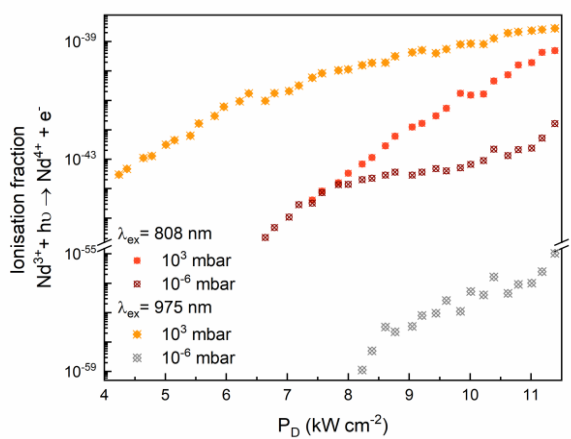

Fig. S16. Ionisation fraction to $Nd^{3+} + h\nu \rightarrow Nd^{4+} + e^-$ at CW-laser excitation on 808 nm and 975 nm, and different pressure conditions, when $E_{ion}$ = 4.6 eV.

## 9. Characterisation of the photoconductivity associated with the LIWE

The error associated with the current density values was determined through the Eq. S7.

$$\Delta J_{Richardson-Dushmann} = J \sqrt{\left(\frac{\Delta A}{A}\right)^2 + \left(\frac{2\Delta T}{T}\right)^2 + \left(\frac{\phi}{k_B T^2} \Delta T\right)^2 + \left(\frac{\Delta \phi}{k_B T}\right)^2}$$

Eq. S7

The photoconductivity resulting from the multiphoton ionisation process produces resistance oscillations on the order of $10^9$ $\Omega$ within approximately $10^{-15}$ seconds. [31–34] In lanthanide materials, photocurrent generation via charge-transfer states is triggered by excitation around 5.9 eV.

## 10. Structural and Morphological modifications of NdMnO₃ after laser irradiation

We analyse $NdMnO_3$ before and after laser irradiation. The XRD pattern and TEM analysis indicate a structural change following laser irradiation at varying pump powers, as previously discussed for other systems exhibiting white light emission characterised by a continuous spectra and obtained under NIR excitation. [35–38] The well-defined diffraction peaks associated with the orthorhombic phase of $NdMnO_3$, as shown in Fig. 1 in the main manuscript, are no longer observed and are instead replaced by a broad halo that corresponds to a lack of crystallinity in the sample after laser irradiation at high power density (on the order of kW·cm⁻²).

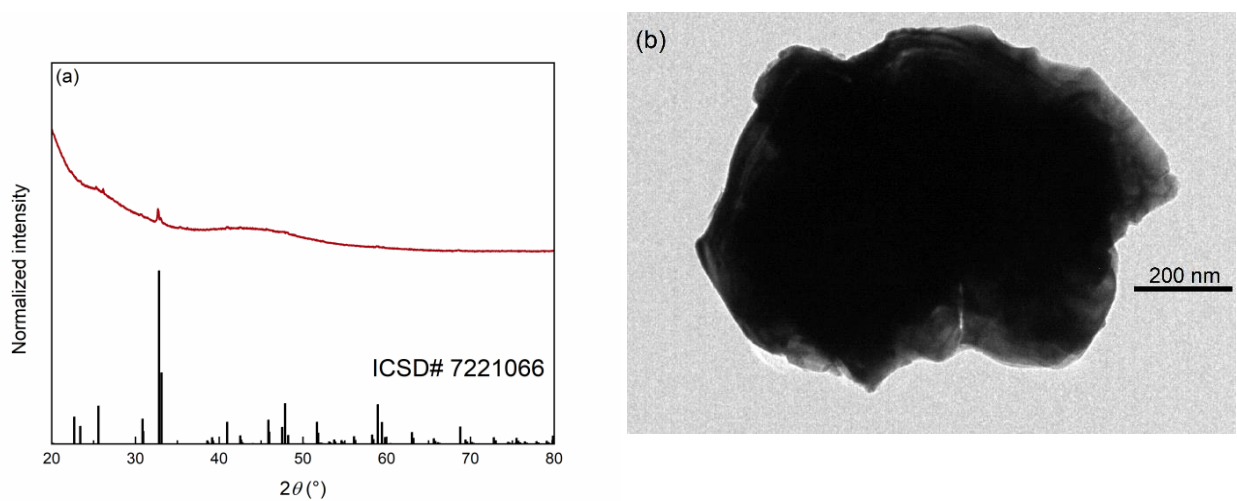

Fig S17. (a) Diffraction pattern and (b)TEM image of the NdMnO3 after laser exposure. XRD acquired in the 2θ range of 20 − 80° at room pressure and temperature (step 0.36°min-1). Phase quantification was obtained using the software X'Pert HighScore

Plus.